\documentclass[useAMS,usenatbib]{mn2e}
\usepackage{graphicx}
\usepackage{amssymb}

\newcommand{\mc}{\multicolumn}

\def\kms{\,{\rm km\,s^{-1}}}
\def\mpc{\,{\rm Mpc}}
\def\kpc{\,{\rm kpc}}

\def\Mbh{\,{M_{\rm BH}}}
\def\msun{\,{M_{\rm \odot}}}
\def\magsec{\,{\rm mag\,arcsec^{-2}}}
\def\Dmin{{\,D_{\rm min}}}
\def\rmin{{\,R_{\rm min}}}

\begin{document}

\title[Photometric Properties and Scaling Relations of BCGs]
{Photometric Properties and Scaling Relations of Early Type Brightest Cluster Galaxies}
\author[F. S. Liu et al.]{
F. S. Liu$^{1,2,3}$\thanks{E-mail: lfs@bao.ac.cn},
X. Y. Xia$^{3}$,
Shude Mao$^{4,3}$,
Hong Wu$^{1}$,
Z. G. Deng$^{2}$\\
$^{1}$National Astronomical Observatories, Chinese
Academy of Sciences, A20 Datun Road, Beijing, 100012, P.R.China\\
$^{2}$Graduate School of the Chinese Academy of Sciences,
                 Beijing, 100049, P.R.China\\
$^{3}$Department of Physics, Tianjin Normal University,
Tianjin, 300074, P.R.China\\
$^{4}$Jodrell Bank Centre for Astrophysics, Alan Turing Building, University of Manchester,
Manchester, M13 9PL, UK
}

\date{Accepted 23 November 2007, Received 23 November 2007; in original form of 6 August 2007 
}

\pagerange{\pageref{firstpage}--\pageref{lastpage}} \pubyear{2007}

\maketitle

\label{firstpage}

\begin{abstract}
We investigate the photometric properties of the early type
Brightest Cluster Galaxies (BCGs) using a carefully selected
sample of 85 BCGs from the C4 cluster catalogue with redshift less
than 0.1. We perform accurate background subtractions, and surface
photometry for these BCGs to 25 $\magsec$ in the Sloan $r$-band. By quantitatively
analysing the gradient of the Petrosian profiles of BCGs, we find
that a large fraction of BCGs have extended stellar envelopes in
their outskirts; more luminous BCGs tend to have more extended
stellar halos that are likely connected with mergers. A comparison
sample of elliptical galaxies was chosen with similar apparent
magnitude and redshift ranges, for which the same photometric 
analysis procedure is applied. We find that BCGs have steeper
size-luminosity ($R \propto L^\alpha$) and Faber-Jackson ($L
\propto \sigma^\beta$) relations than the bulk of early type galaxies.
Furthermore, the power-law indices ($\alpha$ and
$\beta$) in these relations increase as the isophotal limits
become deeper. For isophotal limits from 22 to 25 $\magsec$, BCGs
are usually larger than the bulk of early type galaxies,
and a large fraction ($\sim 49\%$) of BCGs
have disky isophotal shapes. The differences in the scaling
relations are consistent with a scenario where the dynamical structure and formation route
of BCGs may be different from the bulk of early type galaxies, in
particular dry (dissipationless) mergers may play a more important
role in their formation; we highlight several possible dry merger
candidates in our sample.

\end{abstract}

\section{INTRODUCTION}

The Brightest Cluster Galaxies (BCGs) are the most luminous and
most massive galaxies in the universe. BCGs are located close to
the centre of the clusters of galaxies based on the X-ray
observations or gravitational lensing observations (e.g., Jones \&
Forman 1984; Smith et al. 2005). It was noted very early on that
some BCGs show an excess of light (`envelopes') over the de
Vaucouleurs ($r^{1/4}$) profile at large radii (Matthews et al.
1964; Oemler 1973, 1976; Schombert 1986, 1987, 1988; Graham et al.
1996), and a large fraction of BCGs are termed as cD galaxies
(e.g., Patel et al. 2006). Although the origin of such extended
envelopes is still not completely clear (e.g., Patel et al. 2006), the extended
stellar halos of BCGs to surface brightness $\mu(r)<25 \magsec$
are likely from BCGs themselves: the intra-cluster light
has much lower surface brightness and only dominates at large
radius ($r \ga  80 \kpc$; for detailed discussions, see
Zibetti et al. 2005; Bernardi et al. 2007; Lauer et al. 2007).

There has been debate on the formation mechanisms for BCGs and cD
galaxies. Several mechanisms have been proposed, such as galactic
cannibalism (the merging or capture of cluster satellites due to
dynamical friction, Ostriker \& Tremaine 1975; White 1976;
Ostriker \& Hausman 1977), tidal stripping from cluster galaxies
(Gallagher \& Ostriker 1972; Richstone 1975, 1976; Merritt 1985),
and star formation on BCGs by cooling flows (e.g., Fabian 1994).
Therefore, it is important to study the photometric properties of
BCGs using large and homogeneous BCG samples, which has now become
available due to large surveys such as the Sloan Digital Sky
Survey (SDSS; for earlier studies, e.g. Postman \& Lauer 1995, see references
below). Moreover, the statistical properties of the
extended envelopes of BCGs may also help to understand how the
BCGs form and evolve, especially when combined with high-redshift
cluster samples such as EdisCS (White et al. 2005).

Recently, there has been much progress in this area from both
observational and theoretical fronts. Observations show that BCGs
have distinct properties from the bulk of elliptical galaxies. The
size-luminosity relation and Faber-Jackson relation ($L-\sigma$
relation) for BCGs are significantly steeper than those of non-BCG
elliptical galaxies (Lauer et al. 2007; Bernardi et al. 2007;
Desroches et al. 2007; von der Linden et al. 2007). It implies that BCGs are larger and have
lower velocity dispersions than non-BCG elliptical galaxies,
which confirms the pioneering work based on a small BCG sample by
Oegerle \& Hoessel (1991). The differences between BCGs and
non-BCG elliptical galaxies demonstrate that the BCGs may form in a
qualitatively different way from non-BCG elliptical galaxies (De
Lucia \& Blaizot 2007). Furthermore, the steeper $L-\sigma$
relation for BCGs than those of non-BCG elliptical galaxies leads
to a contradiction in the prediction of black hole (BH) masses
using the $\Mbh-\sigma$ or $\Mbh-L$ relationships. The usual $\Mbh-\sigma$
relation cannot predict black hole masses larger than $3 \times 10^{9}$ $\msun$ in the universe, 
which is inconsistent with values estimated by other means (Lauer et al. 2007).

Recent simulations and semi-analytic works in
the cold dark matter hierarchical structure formation framework 
provide a plausible picture for the formation of BCGs: they tend
to form at high-density peaks when their inhabited dark matter
halos collapse at high redshift, and then X-ray driven cooling
flows allow a rapid collapse and the formation of a stellar
component (Cowie \& Binney 1977; Fabian \& Nulsen
1977; Fabian 1994). While more than half of the stellar mass in
the BCGs may have formed before redshift three, BCGs still grow
substantially through dry (dissipationless) mergers when their
host massive halos accrete and merge with other halos since
redshift one (Gao et al. 2004; De Lucia \& Blaizot 2007). This
picture is largely consistent with observations, which does not
require cooling flows to provide the cold gas if BCGs form late;
it also overcomes a potential problem that the merger rate in
clusters may be too low due to the high velocity dispersion in
dynamically relaxed clusters. Furthermore, dry mergers have been
directly observed in cluster environments as well as in the field
(e.g., Lauer 1988; van Dokkum 2005; Tran et al.
2005; Bell et al. 2006). However, there are still no systematic
investigations on the formation of cD-like BCGs galaxies,
especially how the extended stellar halos of cD galaxies form:
whether they appear during the dry merger phase, and whether
they are the primary cause of the distinct properties of BCGs.
There is also no agreement about the contributions of the
intra-cluster light (e.g., Gonzalez et al. 2005; Zebetti et al.
2005; Lauer et al. 2007).

The C4 cluster catalogue (Miller et al. 2005) with 748 cluster of
galaxies was constructed from the spectroscopic data of Second
Data Release (DR2) of the Sloan Digital Sky Survey (SDSS). This is
one of the largest published homogeneous database of nearby
clusters and groups ($z<0.17$). In principle, it is suitable for
statistical studies of BCGs (see also the more recent, much
larger maxBCG catalogue of 13,823 clusters of Koester et al. 2007).
Based on BCGs from this published cluster catalogue, or
the un-published C4 cluster catalogue selected from SDSS DR3,
Bernardi et al. (2007) and von der Linden et al. (2007) find that
BCGs have different properties, in particular their scaling laws,
from non-BCG early type galaxies. Their results are, however, not
entirely consistent with each other, which could be due to
different sample selections or photometry methods. In fact, the
slope indices in the size-luminosity and Faber-Jackson ($L-\sigma$) relations
found by Bernardi et al. (2007) are also significantly different from 
those of Lauer et al. (2007), although the trends are the same and
these two groups use  similar methods to measure luminosities and
effective radius $R_{e}$; both are based on fitting the $r^{1/4}$ law
to the $R-$band surface brightness profiles within 50\kpc\ or to some 
surface brightness limit of the BCGs. However, the
samples they used are different. Lauer et al. (2007) use their own
all sky, volume limited (z$<$0.05) survey of 119 BCGs with
precise ground-based surface photometry (Postman \& Lauer 1995). 
Therefore, it is worth investigating whether sample selections and photometry can lead
to inconsistencies on scaling laws.
In this paper, we construct a different, nearby (with redshift
$z<0.1$) BCG sample based on the C4 cluster catalogue, and perform
our own photometry. In particular, we examine how the scaling
relations change when different isophotal limits are applied, and
compare the results with those of Bernardi et al. (2007), Lauer et
al. (2007) and von der Linden et al. (2007).

The outline of the paper is as follows. In \S2 and \S3 we describe
the selection of our local BCG sample and photometric data
reduction. We present our main results in \S4 and finish with a
discussion \& summary in \S5.
Throughout this paper we adopt a cosmology with
a matter density parameter $\Omega_{\rm m}=0.3$, a cosmological constant
$\Omega_{\rm \Lambda}=0.7$ and
a Hubble constant of ${\rm H}_{\rm 0}=70\,{\rm km \, s^{-1} Mpc^{-1}}$
($h \equiv H_0/100\,{\rm km \, s^{-1} Mpc^{-1}}=0.7$).

\section{SAMPLE} \label{sec:sample}

Our early type BCG sample is drawn from the early version of the
C4 cluster catalogue (Miller et al. 2005). Given that BCGs are
located near the centre of clusters of galaxies which is often
crowded with galaxies and contaminated by intra-cluster light, to
ensure the reliable photometric measurement with high enough S$/$N
ratio for each sample BCG to $\mu(r)$$< 25$ mag/arcsec$^2$ that is
significantly higher than the intra-cluster light (Zibetti et al.
2005; Lauer et al. 2007; Bernardi et al. 2007), we restrict our
BCG sample to be brighter than 14.5 mag in the SDSS $r$-band model
magnitude. After we exclude 10 duplicated objects (Bernardi et al.
2007), there are 114 galaxies selected out of a total of 748 BCGs
in the C4 cluster catalogue.

Due to fiber collisions in the spectroscopic data, it is necessary
to incorporate the SDSS photometric catalogue to avoid missing
BCGs using the C4 algorithm. The DR2 version of the C4 catalogue
tried to correct for this by selecting the brightest cluster galaxy
based on the photometric catalog. However, as several works (e.g.,
Bernardi et al. 2007; von der Linden et al. 2007) noticed, some
stars or spiral galaxies (some of these are located at the edge of
clusters) have been mis-classified as BCGs. we thus performed
visual inspections of two-colour images in the $g$-band and
$r$-band for all the 114 BCG candidates and their 
clusters in a region of $200\kpc \times 200\kpc$ and $2\mpc \times
2\mpc$, respectively. We found that C4 1024, C4 2092, C4 2097, C4
2031, C4 2109, C4 3021 and C4 3268 are seriously contaminated by
adjacent bright stars or foreground galaxies; C4 3047 is located
at the edge of the corrected frame (Stoughton et al. 2002); C4
3235 is a star; C4 2178 and C4 3257 are nearly edge-on galaxies
with dust lanes. In addition, there are 5 objects (C4 1265, C4
1332, C4 2130, C4 3098, and C4 3205) whose parent cluster
richness, defined as the number of galaxies within 1 $h^{-1}\mpc$,
is less than 10, and hence they may be wrong
identifications. In the remaining 98 objects, there are 10 BCGs
that are late type galaxies with obvious spiral arms (C4 1023, C4
1053, C4 1324, C4 1366, C4 2001, C4 3016, C4 3143, C4 3246, C4
3282, and C4 3286). These are also excluded from our early type BCG
sample, leaving us with 88 possible early-type BCGs. We further
excluded 3 objects (C4 1048, C4 1186 and C4 3285) with $M_{r, 25}
> -21$ that are fainter than $L^{\ast}$ galaxies ($M^{\ast}_r \sim
-21.12$ for $h=0.7$, see Blanton et al. 2003a) and they are
unlikely BCGs (Bernardi et al. 2007). The detailed information for
all 29 rejected objects are listed in Table 1. Our final sample
consists of 85 BCGs that are in the redshift mange of
0.03$\leq$z$\leq$0.09 and $r-$band (model) apparent magnitude in
$\sim 13.5< {\rm mag} \leq 14.5$. 

Furthermore, we carefully searched for the second ranked cluster
galaxies in the 85 C4 clusters and found that the BCGs are about 1.6 mag brighter on
average than the second ranked cluster galaxies for the $r-$band absolute magnitude 
at $25\magsec$. Only 6 BCGs have absolute magnitudes within 0.5 mag of the second ranked cluster galaxies.
During this process, we found one mis-identified BCG for C4 2020 (Abell 0119), the brightest galaxy
in this cluster is UGC 00579, instead of ARK 021 as listed in the C4
catalog. We caution that such cases may also exist in the
unpublished C4 catalogue based on the DR3 of the SDSS 
(see von der Linden et al. 2007).

Notice that there are 4 BCGs (C4 2049, C4 1176, C4 3311 and C4
1035) in our BCG sample described above with obvious merger
features. C4 2049, 1176 and 1035 are merger galaxies with two
close nuclei and C4 3311 is with clear asymmetrical shapes. We
will discuss these four objects in more detail in
\S\ref{sec:merger}.

\begin{figure}
\centering
\includegraphics[width = 8cm]{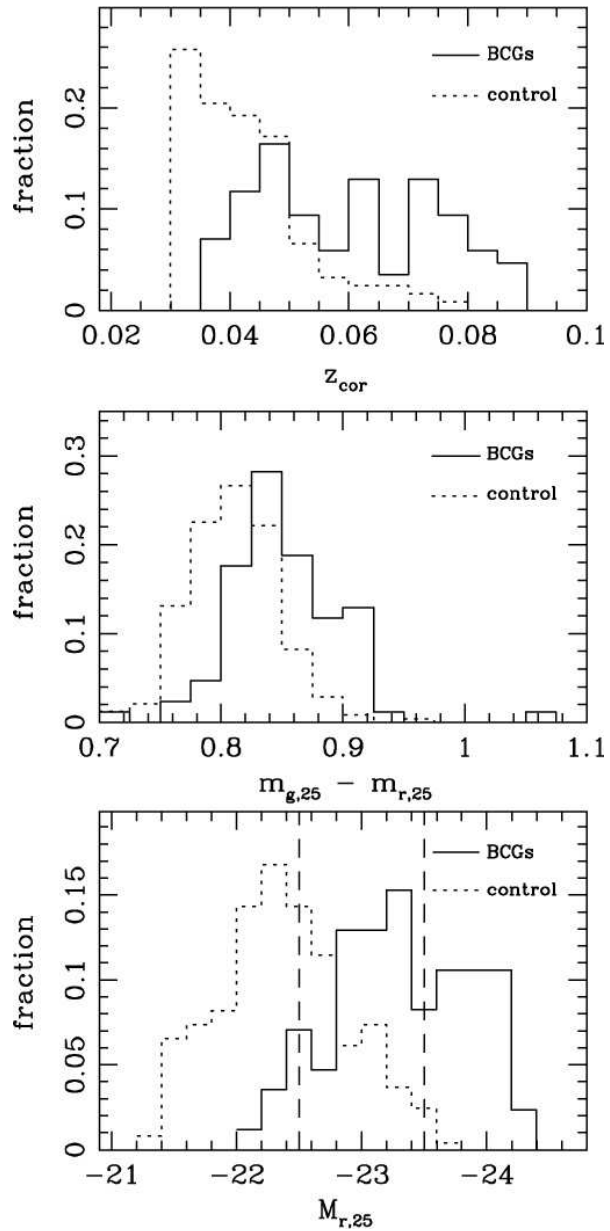}
\caption{Distributions of corrected redshifts (top panel),
(m$_{g,25}- m_{r,25}$) colour (middle panel), and absolute
magnitudes at $25\magsec$ (bottom panel) in the $r$-band for 85 sample BCGs
(solid) and 244 control galaxies (dotted) respectively. Two
vertical dashed lines in bottom panel show $-23.5 \leq M_{r,25}
\leq -22.5$, where the two samples have a significant overlap. We
will compare scaling relations for these two subsamples in
\S\ref{sec:scaling}. } \label{redz_col.eps}
\end{figure}

In order to perform direct comparisons with non-BCGs, we construct
a control sample of normal early-type galaxies from the SDSS DR2
spectroscopic catalogue with galaxy velocity dispersion
measurement in the MPA {\tt gal$\_$info} catalogue
\footnote{http://www.mpa-garching.mpg.de/SDSS/DR4/raw$\_$data.html}.
These control galaxies are chosen to be in the same redshift and
apparent magnitude ranges as sample BCGs, i.e.,
0.03$\leq$z$\leq$0.09 and 13.5$\leq$mag$\leq$14.5. 244 early-type
galaxies are selected as the control sample after excluding nearly
edge-on objects (most of them have dust features), contaminated
sources, and objects identified as C4 BCGs. The distributions of
the corrected redshifts relative to the Local Group (see Blanton
et al. 2005), (m$_{g,25}- m_{r,25}$) colour, and absolute
magnitudes at $25\magsec$ for these two samples are shown in
Fig.~\ref{redz_col.eps}. It can be seen that the control sample
has a lower redshift range than those of BCGs. This is because
non-BCGs are intrinsically fainter than the BCGs\footnote{
Bernardi et al. (2007) found that all galaxies brighter than $-$24 in the $r$-band are
BCGs. Lin $\&$ Mohr (2004) found that BCGs are about one magnitude (0.83) brighter on
average than the second ranked cluster galaxies even within a 
small metric radius of $13.7h^{-1}$ kpc. This does not contradict
with the larger difference (1.6 mag on average) we found for our BCG sample because we measure the magnitudes to 
$25\magsec$, which corresponds to a larger physical size than the radius used by them.
}
(see the bottom panel), so for the same magnitude limit, they reside more locally.
Therefore, it is difficult to construct a control sample in the same
luminosity range as that of BCGs. Instead we attempt a
comparison in a narrow luminosity range,
$-23.5 \leq M_{r,25} \leq -22.5$ (indicated by two vertical
dashed lines at the bottom panel of Fig.~\ref{redz_col.eps}) where 
there are sufficient overlaps between the control sample and BCGs (see
\S\ref{sec:scaling} for comparisons).
Furthermore, both samples have colour $(m_{g,25}- m_{r, 25})\ge
0.7$, implying that all of them are early type galaxies
colour-wise (see Blanton et al. 2003b; Shen et al. 2003 and
reference therein). Also the concentration
index $C=R_{90}/R_{50}$ for virtually all our
galaxies is larger than 2.6, again consistent with their being early-type
galaxies (Shimasaku et al. 2001; Strateva et al. 2001).
All the basic parameters for these 85 BCGs are listed in Table 2.

\begin{table*}
\caption[]{Reasons for the rejection of 29 out of 114 C4 BCGs.}
\begin{center}
\begin{tabular}{cccl}
\hline
\mc{1}{c}{C4 ID} & \mc{1}{l}{R.A.(J2000)} & \mc{1}{l}{Dec.(J2000)} & \mc{1}{c}{reason for rejection} \\
\mc{1}{c}{(1)} & \mc{1}{c}{(2)} & \mc{1}{c}{(3)} & \mc{1}{c}{(4)} \\

\hline
1023& 153.664827&  -0.830910&  late type \\
1053& 228.307961&  4.287565 &  late type \\
1324& 223.045968&  -0.256108&  late type \\
1366& 223.396096&  0.010394 &  late type \\
2001& 350.862674&  14.325841&  late type \\
3016& 187.452774&  64.033013&  late type \\
3143& 247.104069&  41.168463&  late type \\
3246& 136.572283&  50.089290&  late type \\
3282& 249.318427&  44.418227&  late type \\
3286& 184.232511&  63.409922&  late type \\
1024& 226.688031&  -1.231709&  contaminated by bright star \\
2031& 324.773334&  -0.706191&  contaminated by bright star \\
2092& 24.225138 & -0.533071 &  contaminated by bright star \\
2097& 338.797166&  -1.049529&  contaminated by bright star \\
2109& 24.314061 & -9.197611 &  contaminated by bright star \\
3268& 239.339150&  54.671105&  contaminated by bright star \\
3021& 247.489689&  40.630707&  contaminated by foreground spiral \\
2178& 351.297310&  15.199853&  edge-on galaxy with dust lane \\
3257& 249.074493&  44.135700&  edge-on galaxy with dust lane \\
3235& 119.610850&  37.732302&  a star \\
3047& 158.245415&  56.748148&  at the edge of the frame \\
1265& 168.680626&  4.024536 &  richness $<$ 10 \\
1332& 167.820450&  -0.824176&  richness $<$ 10 \\
2130& 50.701678 & -6.694635 &  richness $<$ 10 \\
3098& 246.907102&  42.638233&  richness $<$ 10 \\
3205& 135.367430&  55.044549&  richness $<$ 10 \\
1048& 147.895859&  1.112059 &  $L < L^{\ast}$ \\
1186& 183.990581&  3.305910 &  $L < L^{\ast}$ \\
3285& 259.895020&  56.630066&  $L < L^{\ast}$ \\

\hline
\end{tabular}
\end{center}
\end{table*}

\begin{figure*}
\begin{center}
\includegraphics[angle=0,width=0.71\textwidth]{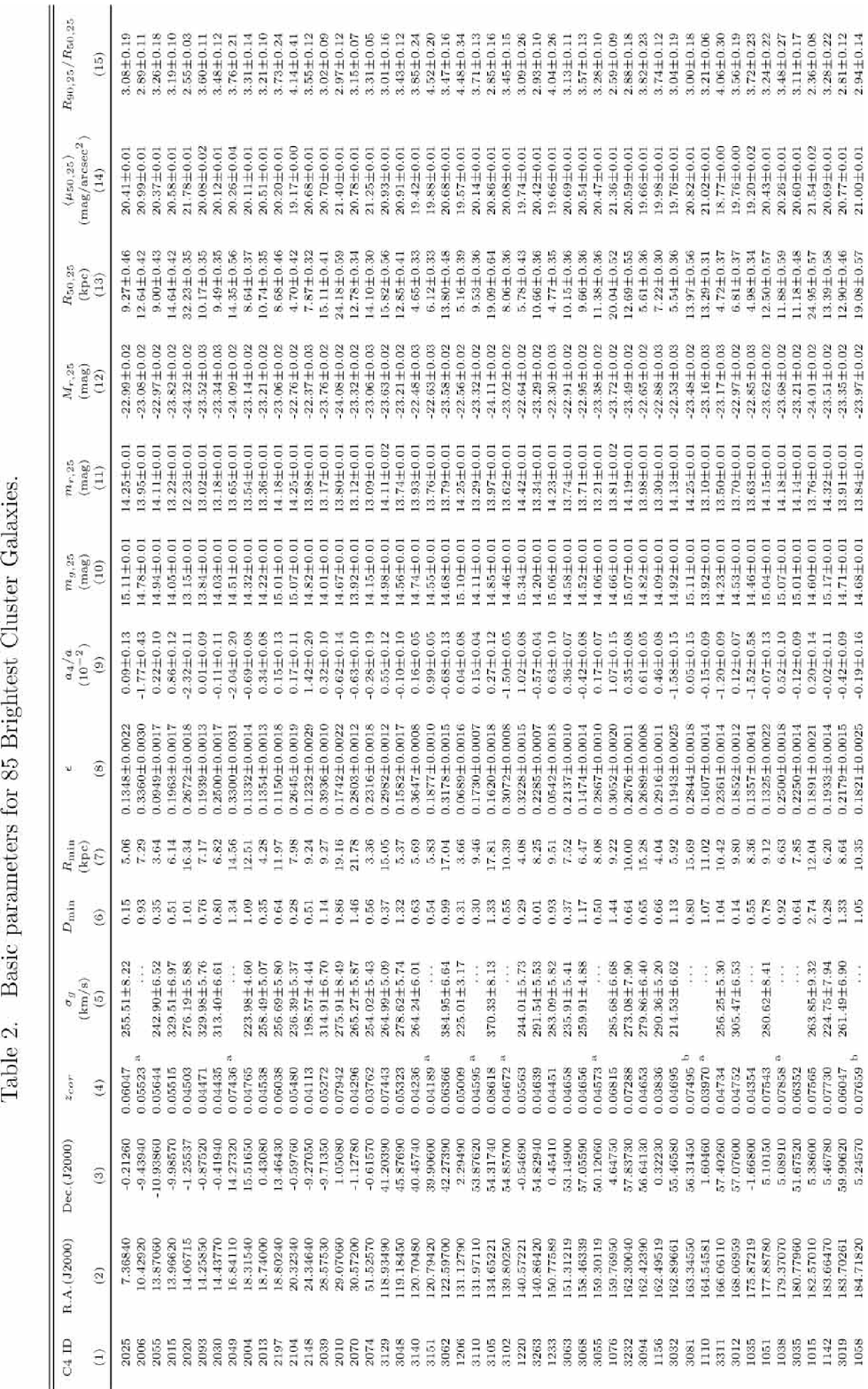}
\end{center}
\end{figure*}

\begin{figure*}
\begin{center}
\includegraphics[angle=0,width=0.65\textwidth]{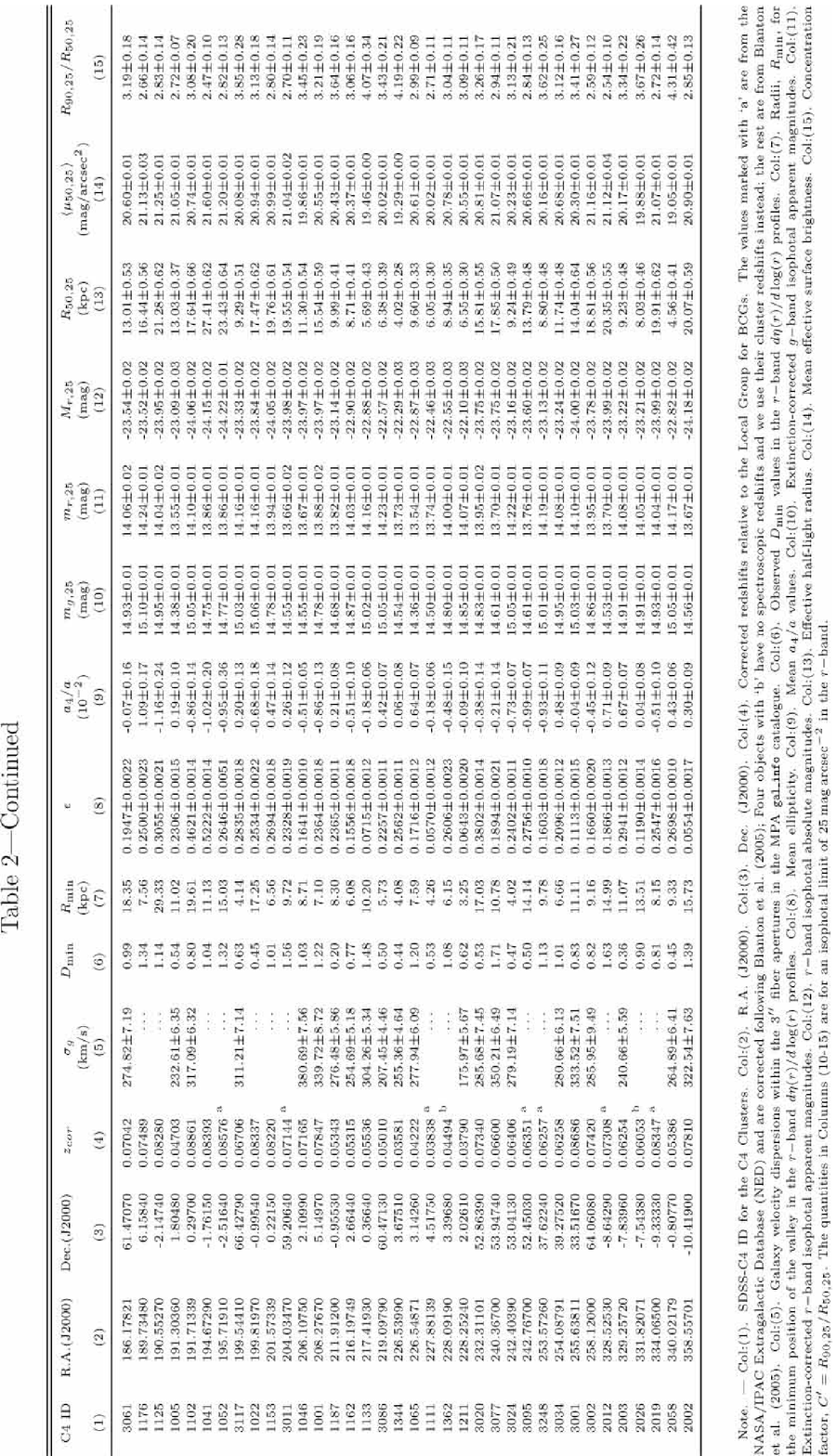}
\end{center}
\end{figure*}

\section{DATA REDUCTION} \label{sec:data}

\subsection{Estimation of the Sky Background} \label{sec:image}

Given that the SDSS image survey in the $u-$ and $z-$bands are
relatively shallow and the $i-$band images suffer from the `red
halo' effect (e.g., Michard 2002; Wu et al. 2005), we use the $g-$
and $r-$band images in this work. The corrected frames have
already been processed by the SDSS photometric pipeline ({\tt
PHOTO}), including bias-subtraction, flat-fielding, cosmic ray
removal and corrections for pixel defects. We use the astrometry
for these two bands obtained by the SDSS astrometry pipeline ({\tt
ASTROM}), which has a typical error of less than $0.1^{''}$
(Stoughton et al. 2002). On the other hand, for reasons we will
discuss below we perform our own photometry.

Although the images from the SDSS archive were processed with
photometric corrections, as Wu et al. (2005) pointed out, some
spurious features still exist in some images. The spatial
variation of these features was about 1-2 ADU. Without
corrections, target galaxies that happen to be located within such
features will have inaccurate background subtraction and poor
surface brightness profiles, especially in the outskirts of the
images. We thus first correct for these structures following Wu et
al. (2005) before constructing the sky background model below.

\begin{figure}
\centering
\includegraphics[width = 8cm]{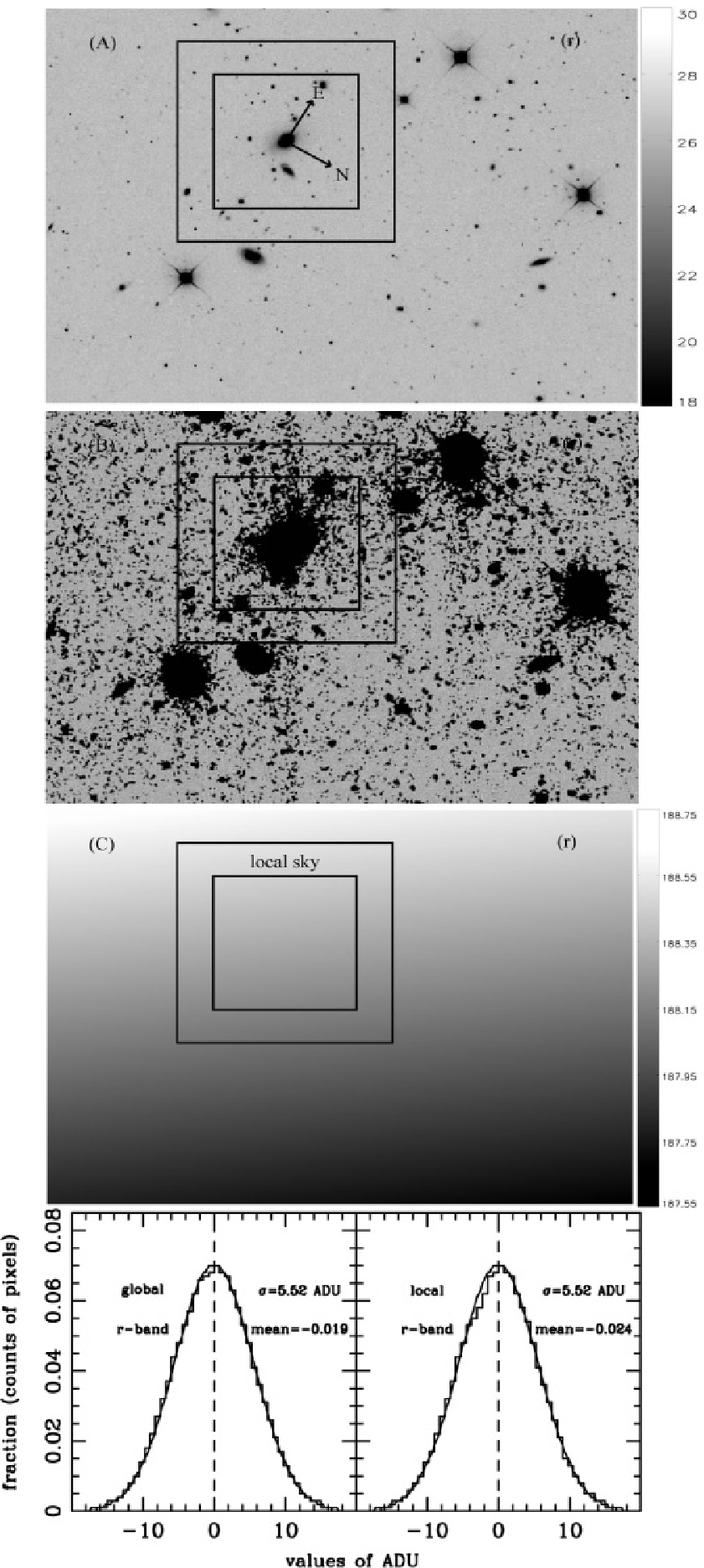}
\caption{ An example of sky subtraction (for target BCG C4 3086)
in the $r-$band. The first row shows the SDSS corrected frame
($2048 \times 1498$ pixels; $13.5\arcmin \times 9.8\arcmin$), the
second row shows the masked frame, and the third row shows the
fitted sky frame. The inner box in each frame has $501 \times 501$
pixels ($3.\arcmin3 \times 3.\arcmin3 $), and the outer box has
$751 \times 751$ pixels ($4.\arcmin7 \times 4.\arcmin7$). The
bottom panels show the normalised distributions of flux in the
unmasked region after the sky background subtraction, for the
whole frame (left) and the local region between the two boxes
(right) in the $r-$band. The solid lines in the two panels are the
best-fit Gaussians to the distributions. }
\label{sky_estimate.eps}
\end{figure}

As Lauer et al. (2007) pointed out that the SDSS photometric 
reduction systematically under-estimates the
luminosities and half-light radius of BCGs. This arises because
the {\tt PHOTO} pipeline often over-estimates the sky background
for galaxies with large size and/or in crowded fields. Both
problems are present for BCGs. As a result, Bernardi et al. (2007)
performed their own photometric reductions for their sample.
On the other hand, von der Linden et al. (2007) address this
problem by adding up to 70\% of the difference between the local
(256 $\times$ 256 pixels) and the global (2048 $\times$ 1498
pixels) sky background (both are available from {\tt PHOTO})
around the BCGs to the radial surface brightness profiles and the
photometry.
However, the sky background in the SDSS frames often shows spatial
gradients and asymmetry (see the third row of
Fig.~\ref{sky_estimate.eps} for an example). Therefore, adding a
constant background value to the radial surface brightness
profiles may not yield the most accurate photometry, at least on
one-to-one basis (see Fig. \ref{sbp_comparison.eps} for
comparisons). Therefore, we performed our own photometry following
the method of Zheng et al. (1999) and Wu et al. (2002), which was
developed to perform deep photometry for nearby large spiral
galaxies. This method has been applied to NGC 5907 and NGC 4565,
which achieved accurate photometry down to an isophotal limit of
$\sim 29\magsec$ in the intermediate bands of the BATC system (Fan
et al. 1996).

To obtain accurate information of the sky background, we first
generate a 2048 $\times$ 1498 pixels ($13.5\arcmin \times
9.8\arcmin$) background-only image using {\tt SExtractor} (Bertin
$\&$ Amounts 1996) by masking all the detected objects with counts
above 1$\sigma$ noise (of the whole frame) in a frame smoothed by
a circular Gaussian with a standard deviation $\sigma$ = 3 pixels.
As most of sample BCGs have sizes of a few square arcminutes, there
are sufficient regions in the masked image to determine the sky
background. A median filter with $51 \times 51$ pixels is then
convolved with the unmasked pixels, after which second-order
Legendre polynomials are used to fit both rows and columns
respectively (see Zheng et al. 1999 and Wu et al. 2002). The
fitted Legendre polynomials are then further smoothed using a
circular Gaussian filter with $\sigma$ = 9 pixels to obtain our
final sky background model. For most BCGs, we find that the sky
background is tilted with a spatial variation about 1 $-$ 2 ADU
across the whole frame. We can subtract this model from the frame
to obtain the sky-free image. Our sky subtraction procedure is
similar to that in Gonzalez et al. (2005). Furthermore, the sky
background subtracted not only includes the contributions of the
intra-cluster light (which is at a much lower level), but also
from other astrophysical sources (dust emission etc., see Gonzalez
et al. 2005 for a discussion).

Our procedure is illustrated in Fig.~\ref{sky_estimate.eps} for C4
3086. The first, second and third rows of
Fig.~\ref{sky_estimate.eps} show the frame, the masked frame and
the smoothed sky background in the $r-$band. Notice that the sky
background shows a gradient across the frame.
The bottom panels show the distributions of counts in the
sky-subtracted frame for all unmasked pixels and for the local
vicinity around the target BCG.
If the sky background model is successful, then we expect the
background counts to follow a Gaussian distribution with a mean
close to zero. This is indeed the case, as can be seen from the
bottom panels of Fig.~\ref{sky_estimate.eps}. The distributions in
the $r-$band for the whole image and the local region are well-fit
by Gaussians with dispersion of 5.52\,ADU and a mean value of
about $-2\times 10^{-2}$ ADU.

After background subtraction, we geometrically align the
two-colour frames, and then trim the corrected frames to
501$\times$501 pixels centred on the target. We run {\tt
SExtractor} again on the trimmed frame to generate a {\tt
`SEGMENTATION'} image, which identifies all objects with flags in
the frames. A mask image with all the detected objects except the
galaxy of interest flagged can then be obtained from the {\tt
`SEGMENTATION'} image. We carefully examined all the mask images
in two colours and corrected a few bad images manually to create
good mask images for all galaxies. Photometry is then
performed on the trimmed images with the masked areas excluded
from the reduction; we discuss the details below.

\subsection{Isophotal photometry} \label{sec:photometry}

To compare the scaling relations, such as the size-luminosity
relation and Faber-Jackson relation, of BCGs with those of the
bulk of elliptical galaxies, we need to perform accurate surface
photometry. In this work, we measure the isophotal magnitudes
within a certain radius, instead of magnitudes based on a model
(e.g., a de Vaucouleurs [1948] model or a S$\acute{\rm e}$rsic
[1968] model) as a large fraction of BCGs is not well described by
such simple models. Furthermore, to see how the scaling relations
vary as a function of isophotal limits, we measure the photometric
parameters to four different surface brightness limits, 22, 23,
24, and 25 mag/arcsec$^2$. We emphasize that the photometry for
the control sample has been performed in the same manner as that for BCGs.

The surface photometry analysis is performed following Wu et al.
(2005). We briefly outline  the procedures below; the readers are
referred to that paper for more details.  The {\tt
ISOPHOTE/ELLIPSE} task in IRAF is used to fit each of the trimmed
background-subtracted images with a series of elliptical annuli
from the centre to the outskirts.
The width of annuli is chosen to increase uniformly in logarithmic
steps, with the semi-major axis radius increasing by 10\% between
two adjacent annuli. The size of the annuli in the outer parts is
therefore larger, which suppresses the shot noise in the outer
regions where the signal-to-noise ratio (S/N) is lower.
The ellipticity, position angle, and other quantities are also
fitted simultaneously.
The $r-$band images are first used to define the isophotal annuli,
which are then applied to the $g-$band images. The coordinates of
the photometric peak in both $r-$band and $g-$band are obtained by
the DAOPHOT package and are fixed during fitting.

We estimate the seeing of our galaxies in two colours by
averaging the star profiles in the corresponding frames
respectively. The average values of the FWHMs of the seeings in
both colours are $\sim1.6''$, which are similar to the measurements
by the SDSS collaboration. We integrate the observed surface brightness profiles
directly to estimate the isophotal apparent magnitudes and
half-light radius $R_{50}$, within which half of the integrated
flux of galaxies is contained, under each isophotal limit. The
absolute magnitude ($M$) is derived from the obtained isophotal
apparent magnitude ($m$) by $M=m-5 \log(D_L/10\,{\rm pc})-A-k$,
taking into account the extinction ($A$) in each filter by the
SDSS, and the k-correction (using the {\tt KCORRECT} algorithm of
Blanton \& Roweis 2007). Other photometric parameters, such as
ellipticities, coefficients of lowest order deviations from
perfect ellipse $a_4$ and $b_4$ (see Bender et al. 1988), are
derived following Hao et al. (2006). Notice that the $a_4/a$
parameters are weighted by the intensity between twice the seeing
radius and the radius where the surface brightness is $25\magsec$
in the $r-$band. We discuss the isophotal shapes in Sect.
\ref{sec:isophotalShapes}.

\begin{figure}
\centering
\includegraphics[width = 8.3cm]{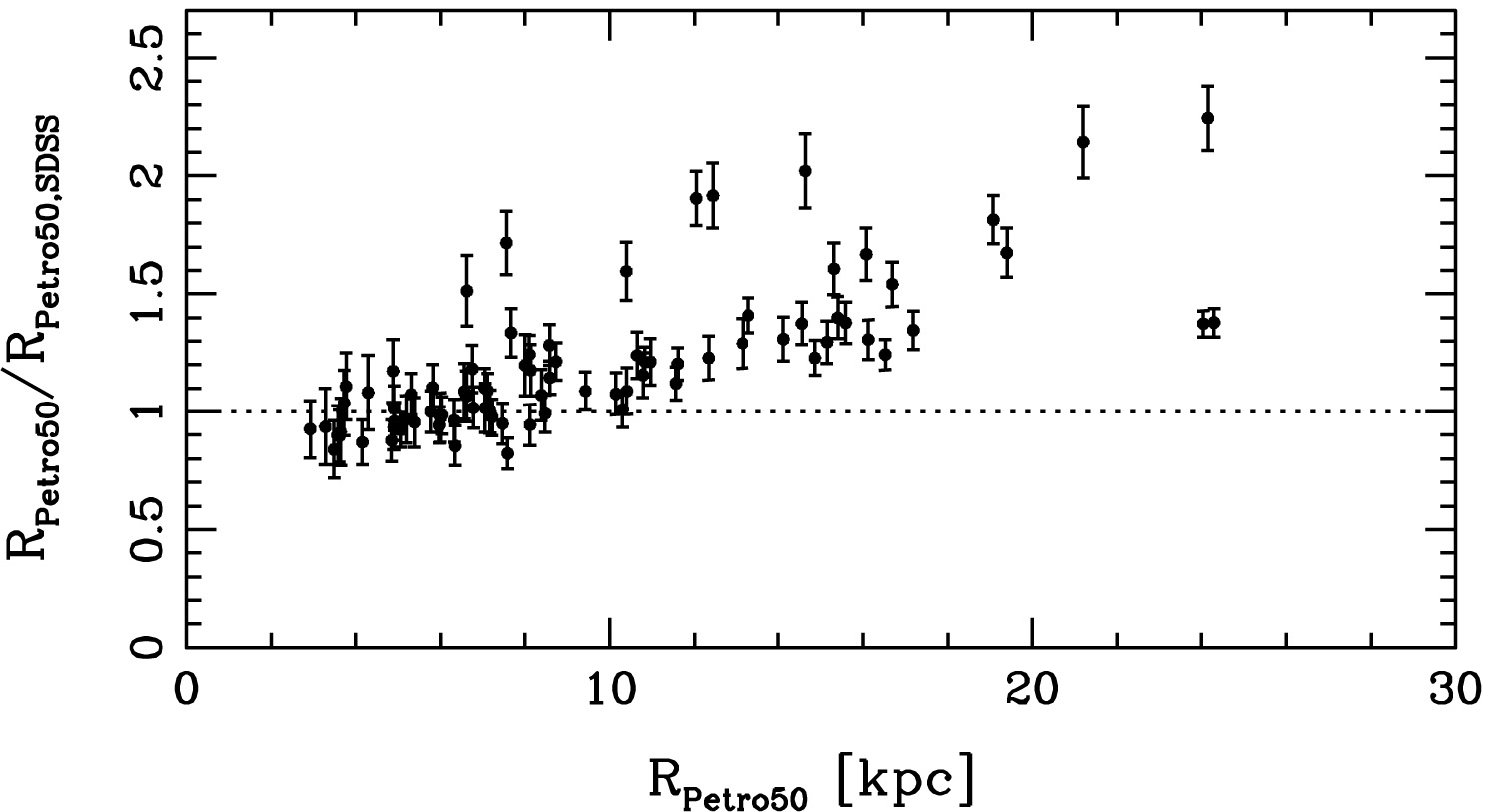}
\caption{
Comparison between our measured Petrosian half-light radii to
those provided by the SDSS for our BCG sample.
}
\label{re_contrast.eps}
\end{figure}

All our radial profiles in this paper use the equivalent radius of an ellipse, $\sqrt{ab}$, where
$a$ and $b$ are the lengths of semi-major and semi-minor axes of the
ellipse. The surface brightness profiles in the two bands
are performed accounting for the Galactic extinction correction, cosmological
dimming, and the k-correction (Blanton \& Roweis 2007). The
photometric errors are estimated as in Wu et al. (2005). The
observational errors in the surface brightness profiles in each
band include random errors (e.g., readout noise, the shot noise of
the sky background and BCGs) and the error from the sky
subtraction.

\begin{figure}
\centering
\includegraphics[width = 8.5cm]{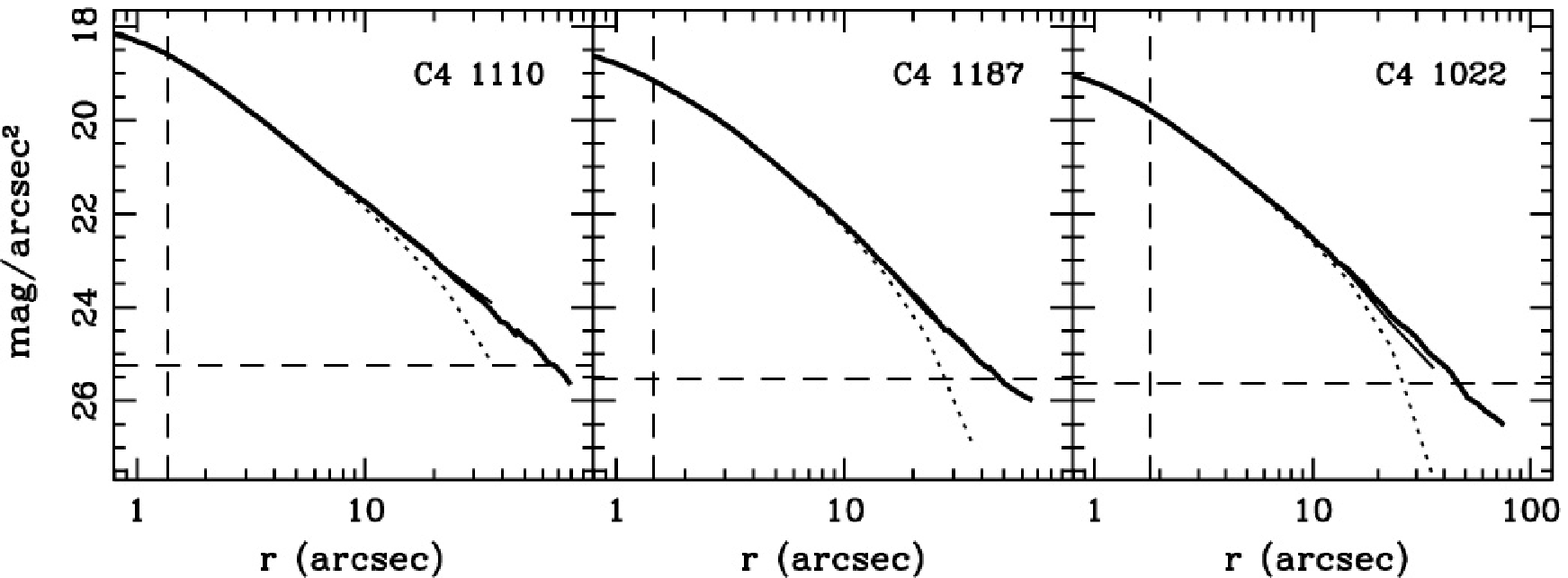}
\caption{ Comparison between the surface brightness profiles
obtained using our method (thick solid lines) and those from the
SDSS (dotted lines) and von der Linden et al. (2007, thin solid
lines) for three example galaxies in the $r-$band. The vertical and
horizontal dashed lines indicate twice seeing radius and one
percent of the sky brightness respectively. Notice for the left
and middle galaxies, our results almost overlap with those from
von der Linden et al. (2007). } \label{sbp_comparison.eps}
\end{figure}

To illustrate the performance of our photometry, we show the comparison
of our determined Petrosian parameters with those of SDSS in
Fig.~\ref{re_contrast.eps}. One can see that the SDSS pipeline
under-estimates the sizes (and thus luminosity) of BCGs, which has
been pointed out by other workers (e.g., Lauer et al. 2007). There is a systematic trend: the
under-estimation becomes more serious for larger (brighter) BCGs.
This is in good agreement with the results of Desroches et al.
(2007, see their Fig. 1.).

As mentioned before, von der Linden et al. (2007) adopted a simple
prescription to correct for the sky subtraction problem. In Fig.
\ref{sbp_comparison.eps} we show a comparison between their
surface brightness profile, ours and that from the SDSS for three
example galaxies.  It can be seen that our results are in very
good agreement with those of von der Linden et al. (2007). While
there are small differences at faint surface brightness
($25\magsec$), such differences have little impact on their
results since they adopted a surface brightness of $23\magsec$ to
measure the physical quantities. So for statistical purposes,
their method can offer an efficient way of performing
photometry, particularly at high surface brightness limits.

\section{RESULTS}

\subsection{The Extended Envelopes of BCGs}

As discussed in the introduction, the surface brightness profiles
of the majority of cD galaxies show strong deviation from a
perfect de Vaucouleurs or S$\acute{\rm e}$rsic profile due to cD
galaxies embedded in an extensive luminous stellar halo. In
order to quantitatively measure the extended envelope of BCGs, we
use an objective method based on the Petrosian $\eta(r)$ profiles
(Patel et al. 2006), defined as
\begin{equation}
\eta(r) \equiv \mu(r) - \langle \mu(r) \rangle,
\label{eq:eta}
\end{equation}
where $\mu(r)$ is the surface brightness in magnitudes at radius
$r$ and $\langle \mu(r) \rangle $ is the mean surface brightness
within $r$ (Petrosian 1976). There is a distinct signature of a
plateau in the Petrosian $\eta(r)$ profiles for cD galaxies with
an extended stellar halo (see below). Such a plateau is not
present for normal elliptical galaxies which are usually well-fit
by a de Vaucouleurs' surface brightness profile 
(e.g., Kj$\ae$rgaard et al. 1993; Brough et al. 2005). Furthermore,
notice that the plateau is not present for the S$\acute{\rm e}$rsic surface
brightness profile that better fits some ellipticals
(see Fig. 1a of Graham et al. 1996).   

\begin{figure}
\centering
\includegraphics[width = 8.4cm]{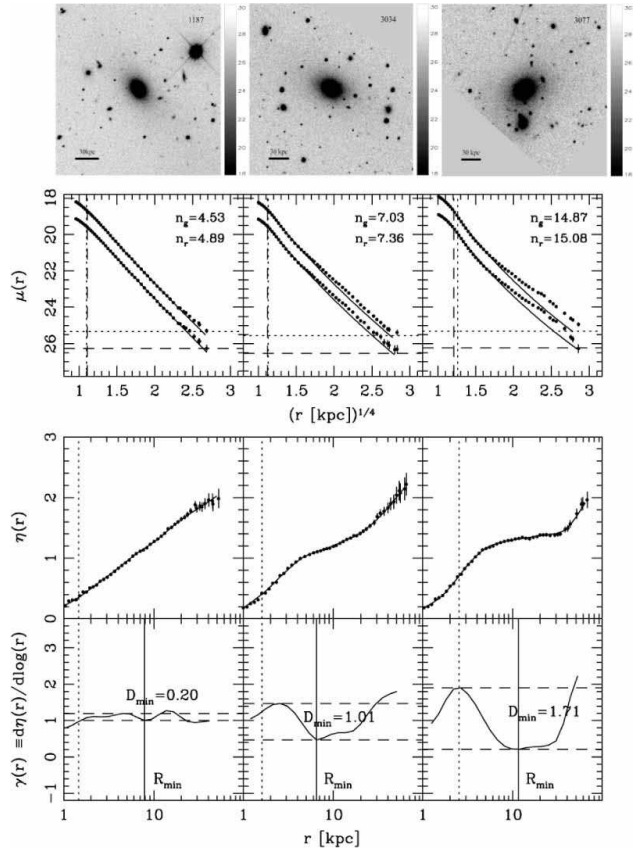}
\caption{Images, surface brightness, $\eta(r)$ and
$d\eta(r)/d\log(r)$ profiles for three typical examples BCGs (C4
1187, C4 3034 and C4 3077). The images shown in the top panels are
in the $r-$band and have a size of $501 \times 501$ pixels. In
each image, north is up and east is to the left. The second row
shows the radial surface brightness profiles in the $g-$band
(lower curve) and $r-$band (upper curve) respectively. The solid
line is the best-fit S$\acute{e}$rsic model for each surface
brightness profile that is obtained by fitting intensity
profile from twice seeing radius to radius at $25\magsec$ respectively. The
vertical and horizontal lines are twice the seeing radius and one
percent of the sky brightness in the $g-$band (dashed lines) and
$r-$band (dotted lines) respectively. The third row shows the
observed (dots) and smoothed (line) Petrosian $\eta(r)$ profiles
in the $r-$band. The bottom row shows the corresponding
$d\eta(r)/d\log(r)$ profile. The vertical solid line in each panel
indicates the minimum position,
}
\label{example_sbp_eta.eps}
\end{figure}

Fig.~\ref{example_sbp_eta.eps} shows the surface brightness
profiles and Petrosian $\eta(r)$ profiles for 3 BCGs. For BCG C4
1187, its profile is well fit by a S$\acute{e}$rsic with $n\approx4.53 $ (in the $g$-band)
model, and the $\eta(r)$ profile does not show any distinct
feature. In contrast, for C4 3034 and C4 3077, their surface
brightness profiles deviate significantly from a single
S$\acute{\rm e}$rsic profile (or the $r^{1/n}$ profile,
$n=4$ for the de Vaucouleur profile), and the $\eta(r)$ profile shows a
plateau. Furthermore, we can see that the more significant the
deviation of the surface brightness profile from the a single
S$\acute{\rm e}$rsic  profile (i.e., the more prominent the
stellar halo in the outskirt), the flatter the plateau in the
$\eta(r)$ profile becomes.

To quantitatively determine how a surface brightness profile
deviates from the $r^{1/n}$ profile we calculate the gradient of
$\eta(r)$, $\gamma(r) \equiv d\eta(r)/d\log(r)$. 
As mentioned before, a single $r^{1/n}$ model profile always 
results in a monotonic $\gamma(r)$ profile for any $n$.
The bottom panels of Fig.~\ref{example_sbp_eta.eps} show this quantity as a function
of radius. It is clear that galaxies with the most significant
deviation from the model profile also have deeper valleys in the
$\gamma(r)$ profile. 
We label the radius of the minimum in
$\gamma(r)$ as $\rmin$, and also define a depth, $\Dmin$, as the
difference between the minimum and maximum of $\gamma(r)$ outside
twice the seeing radius and $\rmin$.

\begin{figure}
\centering
\includegraphics[width = 8.4cm]{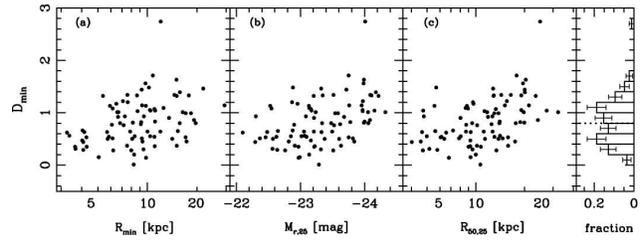}
\caption{
Correlation of $\Dmin$ with $\rmin$, $r-$band absolute magnitude,
and size for 85 BCGs, measured out to an isophotal limit 25 $\magsec$.
The right-most panel shows the histogram of the $\Dmin$ distribution
together with the Poisson errors; the
median value is $\sim 0.8$ (indicated by a thick dotted line).
}
\label{rela1.eps}
\end{figure}

The right histogram in Fig.~\ref{rela1.eps} shows the distribution
of $\Dmin$ for our sample BCGs (the median value is $\sim 0.8$).
We can see that the distribution of $\Dmin$  shows a weak
bi-modality, but is still consistent with a continuous
distribution within Poisson error bars. This shows that the
deviation of the surface brightness profiles from the a single
S$\acute{\rm e}$rsic profile does not have any sharp transitions.
Therefore, it is difficult to give an unambiguous criterion to
separate cD galaxies from non-cD BCGs.

The left panel of Fig.~\ref{rela1.eps} shows the correlations
between $\Dmin$ and $\rmin$. The null-hypothesis that there is no
correlation between $\Dmin$ vs. $\rmin$ can be rejected at the $7.5 \times 10^{-4}$ level. So
statistically the larger the radius $\rmin$, the more significant
the extended envelope a BCG appears to have. The middle and right
panels in Fig.~\ref{rela1.eps} show the correlations between
$\Dmin$ with the $r-$band luminosity and the half-light radius for
BCGs. The null-hypothesis that there are no correlations between
$\Dmin$ vs. $M_{r,25}$, and $\Dmin$ vs. $R_{50,25}$ can be
rejected at the $3.8 \times 10^{-5}$, and  $9.6 \times 10^{-6}$
confidence levels. These correlations demonstrate that more
luminous and larger BCGs tend to have more extended stellar
envelopes and therefore the fraction of cD galaxies increases.

\subsection{Mergers and Envelopes in BCGs \label{sec:merger}}

\begin{figure}
\begin{center}
\begin{minipage}[t]{0.5\linewidth}
\centerline{\includegraphics[angle=0,width=0.99\textwidth]{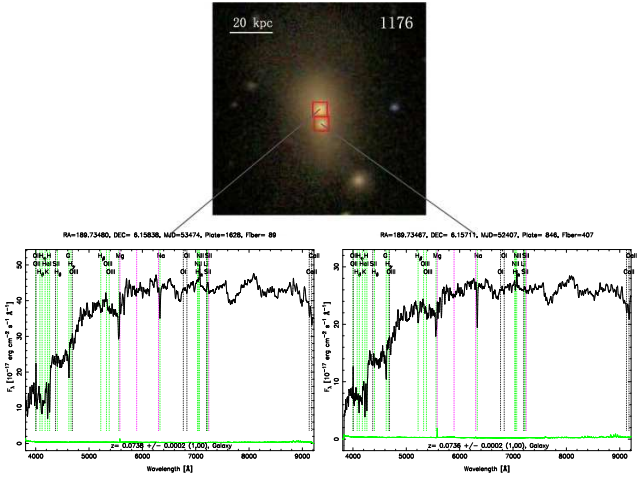}}
\end{minipage}%
\vspace{-4mm}
\\
\begin{minipage}[t]{0.5\linewidth}
\centerline{\includegraphics[angle=-90,width=0.99\textwidth]{f7b.eps}}
\end{minipage}%
\begin{minipage}[t]{0.5\linewidth}
\centerline{\includegraphics[angle=-90,width=0.99\textwidth]{f7c.eps}}
\end{minipage}%
\end{center}
\caption{ The image of merging BCG C4 1176 with two nuclei shown
in the top panel. North is up and east is to the left.  The bottom
two panels show the spectra of the two nuclei (the line symbols
are from SDSS, see http://www.sdss.org/gallery/gal\_spectra.html);
the redshifts of the two nuclei are almost the same, indicating
they are merging. Notice the broad low-surface brightness fan to
the north of the image. } \label{merger_spec.eps}
\end{figure}

As mentioned in \S\ref{sec:sample} that there are four galaxies in
our BCG sample that have either two close nuclei or other clear
merging signatures, such as broad faint fans or asymmetry in the
morphology. The top panel of Fig.~\ref{merger_spec.eps} shows the
image of C4 1176, from which one can clearly see two nuclei and a
broad surrounding fan to the north of the image. The SDSS provides
the spectra for both nuclei, which are shown in the bottom panels
of Fig.~\ref{merger_spec.eps}. From the spectra the redshifts of
the two nuclei are 0.0738 and 0.0736 respectively, corresponding
to only 60 $\kms$ difference in the line-of-sight velocity.
The angular distance between the two nuclei, $\sim 4.7\arcsec$, implies a projected separation of just
$\sim 6.7\kpc$. Therefore, C4 1176 is  most likely a merger galaxy
with two nuclei. Furthermore, the spectroscopic properties of these two nuclei are
similar; both are typical of early type galaxies with weak
emission lines of LINER characteristics,
indicating that BCG C4 1176 may be a dissipationless (dry) merger system.

\begin{figure}
\centering
\includegraphics[width = 8.4cm]{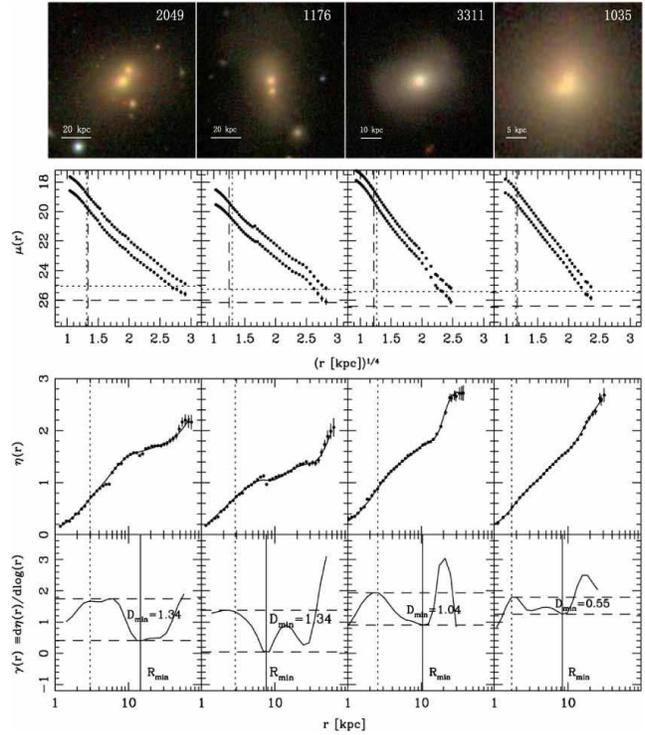}
\caption{ Images, surface brightness, $\eta(r)$  and
$d\eta(r)/d\log(r)$ profiles in the $r-$band for four likely
merger BCGs in our sample. The format is the same as in Fig.
\ref{example_sbp_eta.eps}. In each image, north is up and east is
to the left. } \label{merger_sbp_eta.eps}
\end{figure}

For the other three BCGs, we can see from the top panel of
Fig.~\ref{merger_sbp_eta.eps} that C4 2049 and C4 1035 also have
two possible close nuclei. From the SDSS spectra, we find that
the projected distance between the two nuclei for C4 1035 is only
$\sim 1.7\kpc$ and the relative line-of-sight velocity is 120
$\kms$, and thus may be another merging BCG. For C4 2049, the SDSS
has only spectrum for one nucleus. The angular distance
between these two nuclei is $\sim 5.5\arcsec$. If they are at the
same redshift, then the projected distance is $\sim 7.7 \kpc$.
Hence C4 2049 is another possible merger BCG. For C4 3311,
although one cannot see double nuclei at its centre (down to the
SDSS spatial resolution), it shows clear asymmetry in shape, which
is indicative of merging activities.

The second, third and bottom rows of
Fig.~\ref{merger_sbp_eta.eps} show the surface brightness
profiles, $\eta(r)$ profiles and $d\eta(r)/d\log(r)$ profiles for
all four merger/possible merger BCGs. As can be seen, there are
obvious plateaus in the $\eta(r)$ profiles for 3 out of 4 merger
BCGs (C4 2049, 1176, and 3311). The $\Dmin$ values measured from
the $d\eta(r)/d\log(r)$ profile are 1.34, 1.34, and 1.04, for C4
2049, 1176, and 3311 respectively, which are much larger than the
median $\Dmin$ value ($\sim 0.8$) for the whole BCG sample.
Therefore, these three BCGs clearly have extended envelopes in
their outskirts and are likely cD galaxies.

The high fraction of cD galaxies in merger BCGs and the fans
corresponding to cD galaxy envelopes suggest that the
extended stellar halos of BCGs are from mergers and they
appear shortly after merging takes place.

\begin{figure}
\centering
\includegraphics[width = 8.4cm]{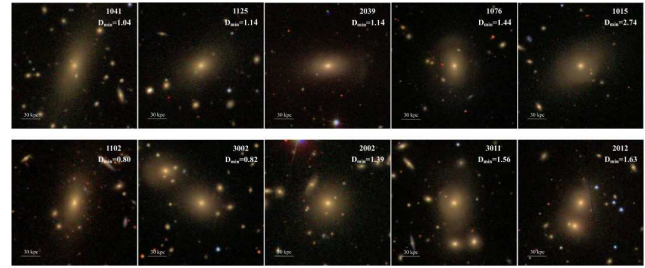}
\caption{ The panels in the top row show the images of BCGs with
obvious dry merger signatures, such as broad fans or faint
asymmetry shapes. In the bottom row, all 5 images of BCGs are
surrounded by many possible satellites or are in pair or triple
systems; these are possible signatures of interaction and mergers.
The values of $\Dmin$ for all these dry merger candidates are
larger than the median value (0.8) in the $\Dmin$ distribution for
sample BCGs. In each image, north is up and east is to the left. }
\label{example_merger.eps}
\end{figure}

In fact, the role of dry (dissipationless) mergers in the
formation of ellipticals has recently received much attention. As pointed
out by Bell et al. (2006) and van Dokkum (2005), the typical
morphological signatures of dry mergers are broad stellar fans,
short tidal tail with similar colour as the galaxy itself, and
asymmetries at very faint surface brightness levels. The four
examples of merger BCG candidates in our sample shown at
Fig.~\ref{merger_sbp_eta.eps} indeed have all
these signatures, but they also show even more direct merger
signs: in three of these four cases, there are two close nuclei.
%
Fig.~\ref{example_merger.eps} shows ten more examples of possible
merger systems from our sample. They have broad low-surface brightness
fans or distinct asymmetries (top panels); many visually
appear to inhabit in environment with multiple small satellites
(bottom panels). These images provide strong evidence for the
connection of mergers with the formation of stellar halos in BCGs.
The 59 BCGs with spectroscopy from SDSS have spectra
similar to the examples shown in Fig.~\ref{merger_spec.eps}, indicating
our BCG samples are typical early-type galaxies.  Their
spectra do not show significant star formation at the centre,
although one third of our BCGs show weak emission lines with
LINER characteristics. Therefore, the mergers appear to be largely
dissipationless (dry). We will return to the role of mergers in the formation of BCGs
in a future study.

\subsection{Isophotal Shapes} \label{sec:isophotalShapes}

We also investigated the shapes of isophotes for BCGs and compared
with those of the control sample. The isophotes of ellipticals are
usually well fit by ellipses. However, small but significant
deviations exist, which are usually described by
the Fourier components of the deviations. The most significant one
is the $a_4/a$ (Lauer et al. 1985; Bender et al. 1989), with a positive (negative)
$a_4/a$ indicating a disky (boxy) isophotal shape.

\begin{figure}
\centering
\includegraphics[width = 9cm]{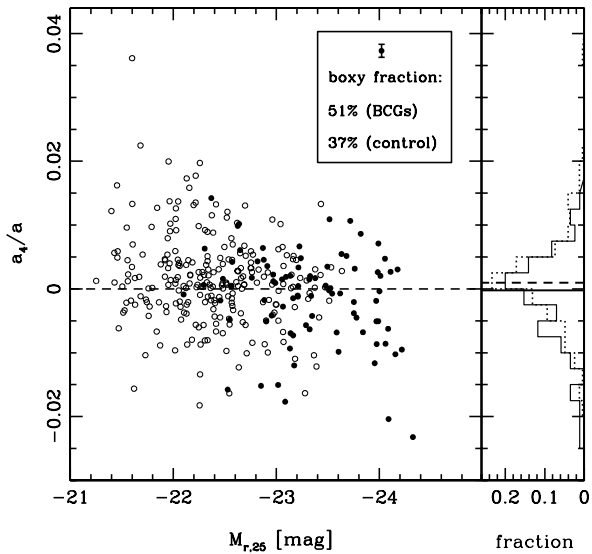}
\caption{ Structure parameter $a_4/a$ versus $r-$band absolute
isophotal magnitude, $M_{r,25}$, for the BCGs (solid circles) and
the control sample (open circles) respectively (left panel). The
median error bar in $a_4/a$, and the fractions of galaxies with
boxy isophotes for BCGs and the control sample are shown in the
inset. The right panel shows the histograms of the $a_4/a$
distributions for the BCGs (solid histogram) and the control
sample (dotted histogram). The median values are shown as solid
and dashed lines respectively. } \label{a4_Mag.eps}
\end{figure}

Fig.~\ref{a4_Mag.eps} shows how the $a_4/a$ parameter varies with
the luminosity of galaxies for both the BCGs and the control
sample. It is clear from Fig.~\ref{a4_Mag.eps} that $a_4/a$
decreases as the luminosity of galaxies increases, hence the
fraction of boxy galaxies increases for both samples; this trend has
been pointed out by many authors (e.g., Bender et al. 1989; Hao et
al. 2006). However, the boxy fraction for BCGs (51\%) is not much
larger than that of the bulk of ellipticals (37\%), even though
BCGs are more luminous than the bulk of ellipticals in the control
sample. In particular, we can see from Table 2 and
Fig.~\ref{a4_Mag.eps} that the boxy fractions of BCGs with
luminosities in the range of $-24 \la {\rm mag} \leq -23$ and
mag$<-24$ are $\sim 55\%$ and $\sim 65\%$ respectively. So there
are still a large fraction of very luminous early type galaxies
with disky isophotes ($a_4/a > 0$), which is consistent with the HST
image results on BCGs by Laine et al. (2003). Faber et al. (1997)
and Lauer et al. (2005) pointed out that galaxies brighter than
$M_V=-22$ mag (corresponding to $M_r \approx -22.6$ mag, after
correcting the difference in $H_0$, and taking $V-r=0.3 \pm0.1$,
Krick et al. 2006) tend to have `cored' luminosity profile,
while  galaxies fainter than $M_V=-20.5$ mag (corresponding to
$M_r \approx -21.1$ mag) tend to have `power-law' luminosity
profiles (see also Rest et al. 2001; Lauer et al. 2005).
Other works (e.g., Rest et al. 2001) show that bright
galaxies are slowly rotating, boxy systems (Davies et al. 1983)
while faint elliptical galaxies are fast rotating, disky systems
(Bender 1988; Bender et al. 1989; Nieto et al. 1988). However,
BCGs appear to somewhat contradict with this trend: some very
luminous BCGs still have disky isophotes. The flattening in
elliptical galaxies can arise either from an-isotropic motions or
from rotation. Faint ellipticals appear to be flattened by
rotation (e.g., van den Bosch et al. 1994; Lauer et al. 2005),
while the luminous ones may be supported by an-isotropic motions.
As we have shown in the last subsection, BCGs may have experienced
fairly recent merging events, and thus may still retain some of the
orbital angular momentum from merging, and their isophotes may be
disky.

\subsection{Scaling relations} \label{sec:scaling}

There have already been many works on the scaling relation of BCGs
based on homogeneous BCG samples (Bernardi et al. 2007; Desroches
et al. 2007; von der Linden et al. 2007; Lauer et al. 2007).
However, so far no work has investigated how the slopes
change with the isophotal limit, which can in principle provide
more knowledge of the dynamical structure of BCGs.

\begin{figure}
\centering
\includegraphics[width = 8.4cm]{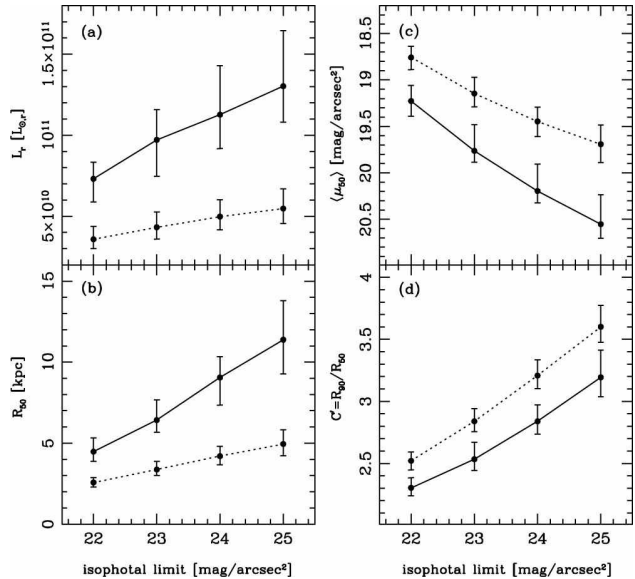}
\caption{ Luminosity ($L_{r}$), size ($R_{50}$), mean surface
brightness within $R_{50}$ ($\langle\mu_{50}\rangle$) and the
concentration factor ($C^{\prime}=R_{90}/R_{50}$) as a function of
the isophotal limit. The solid and dotted lines denote the BCG and
control samples respectively. Error bars are the 68.3$\%$
confidence levels of the median value. } \label{para_contrast.eps}
\end{figure}

\subsubsection{The size-luminosity relation}

The left panels of Fig.~\ref{para_contrast.eps} show how the size
($R_{50}$) and luminosity ($L_{r}$) vary with the isophotal
limits. It is clear that the size ($R_{50}$) and luminosity of
BCGs increase more sharply than that of the control sample as the
isophotal limit becomes deeper. The right panel shows how the mean
surface brightness within $R_{50}$ ($\langle\mu_{50}\rangle$) and
the concentration factor ($C^{\prime}=R_{90}/R_{50}$) vary with
the isophotal limit. We can see 
that $\langle\mu_{50}\rangle$ of BCGs decreases more sharply than
that of control sample, but $C^{\prime}=R_{90}/R_{50}$ of BCGs has
the same trend as that of the control sample as the isophotal
limit becomes deeper. Therefore, BCGs appear to have significantly
larger average sizes and are more diffuse in surface brightness.
These trends agree with those found in previous studies (e.g.,
Lugger 1984; von der Linden et al. 2007).

\begin{figure}
\centering
\includegraphics[width = 8.4cm]{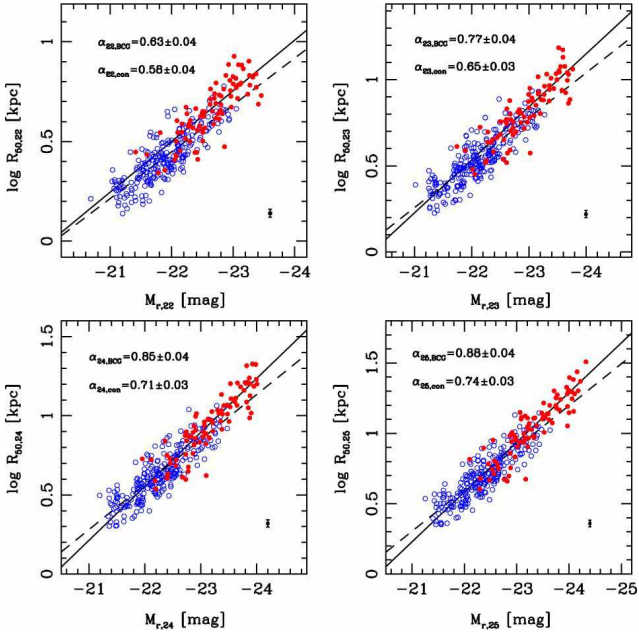}
\caption{ The size-luminosity relation for BCGs (red solid circles) and the
control sample (blue open circles) for four isophotal limits, 22, 23,
24, and 25 $\magsec$ respectively. The median error bar for the
data points is indicated in the bottom right corner of each panel.
The black solid and dashed lines are the best power-law fits for the
BCGs and control sample respectively. The power-law index $\alpha$
($R_{50} {\propto} L^{\alpha}$) is shown at the top of each panel.
} \label{size_L.eps}
\end{figure}

Fig.~\ref{size_L.eps} shows the size-luminosity relations for our
BCGs and the control sample, measured to four isophotal limits,
$22$, $23$, $24$ and $25$ mag/arcsec$^2$ respectively. The solid
and dotted lines in each panel show the best power-law ($R_{50}
\propto L^{\;\alpha}$) fits; the $\alpha$ value is indicated at
the top left corner of each panel.\footnote{Notice that the
intrinsic scatters in the data points are much larger than the
statistical error bars, and so the formal $\chi^2$ per degree of
freedom is large, indicating the presence of systematics. To
approximately account for these, we renormalised the statistical
errors by a constant factor until the $\chi^2$ per degree of
freedom is unity. The errors on the power-law indices are obtained
using these renormalised errors. A similar procedure was adopted
in von der Linden et al. (2007).}

Clearly the size-luminosity relation becomes steeper as the
isophotal limit becomes fainter for both the BCG and the control
sample.
However, we can also see that the size-luminosity relation for
BCGs is steeper than that of the control sample at the same
isophotal limit.
In particular, the size-luminosity relations for BCGs is already
slightly steeper (but not very statistically significant)
than those of the bulk of ellipticals when the
isophotal limit is $22 \magsec$, corresponding to a physical
half-light radius of about 5\,kpc (see Fig.
\ref{para_contrast.eps}).
It indicates that the deviation of size-luminosity relation for
BCGs from that of bulk of ellipticals may already occur for the inner
regions of ellipticals, and thus is not because most BCGs have
extended stellar halos. Rather, this suggests the deviations arise
from different dynamical structures of BCGs from the bulk of
ellipticals. Bernardi et al. (2007) arrived at a similar
conclusion by comparing the size-luminosity relations for BCGs
with and without extended stellar envelopes (see also von der
Linden 2007).

The power-law index ($0.88 \pm 0.04$) for the size-luminosity
relation we derived for BCGs at an isophotal limits of $25$
mag/arcsec$^2$ is slightly smaller than the value ($0.92$)
obtained by Bernardi et al. (2007), but the power-law index
($0.74\pm 0.03$) for the control sample at the same isophotal
limit is a bit larger than their value ($0.62$). The difference in
the slope measured by us and by Bernardi et al. (2007) may be
because we measure the isophotal magnitudes by integrating the
observed surface brightness profiles directly, while Bernardi et
al. (2007) measured the magnitude using a de Vaucouleurs or
S$\acute{e}$rsic model for the surface brightness profile.
In fact, when we use the same method as Bernardi et al. (2007) to
perform photometry, the power-law index for the
size-luminosity relation we obtain is almost the same ($0.93 \pm 0.03$) 
as that of Bernardi et al. (2007), indicating that different
ways of photometry can give different power-law indices in 
the size-luminosity relation.

von der Linden (2007) also derived the size-luminosity relation
down to 23 $\magsec$ for their BCGs. The slope they derived is
$\alpha=0.65\pm 0.02$,  similar to that for their control
sample ($\alpha=0.63\pm 0.02$). Our slope for the BCGs ($0.77\pm
0.04$) is steeper than their value at the same isophotal limit,
while the slope for our control sample ($\alpha=0.65\pm 0.03$)
agrees well with theirs. We have 66 overlapping galaxies with the
BCG samples used by von der Linden et al. (2007). If we use only
these galaxies, we derive a slope similar to theirs. Our steeper
slope appears to arise from the fact that our BCGs are brighter
since their sample includes more fainter BCGs which makes their slope 
shallower (von der Linden 2007, private communication). Therefore, the sample construction
also influences the power-law index in the size-luminosity relation.

The power-law index $\alpha=1.18\pm 0.06$ in the 
size-luminosity relation by Lauer et al. (2007) for their core
galaxies with $M_{V} \le -21$ (most core galaxies with $M_{V} \le
-22$ are BCGs) measured in a similar way as Bernardi et al.
(2007) is significantly steeper than those of Bernardi et al.
(2007), von der Linden (2007) and our results. It again
illustrates that how samples are constructed and the methods used to do
the photometry lead to different power-law indices in the size-luminosity relation.
Comparisons between different studies should be aware of this.

\begin{figure}
\centering
\includegraphics[width = 8.4cm]{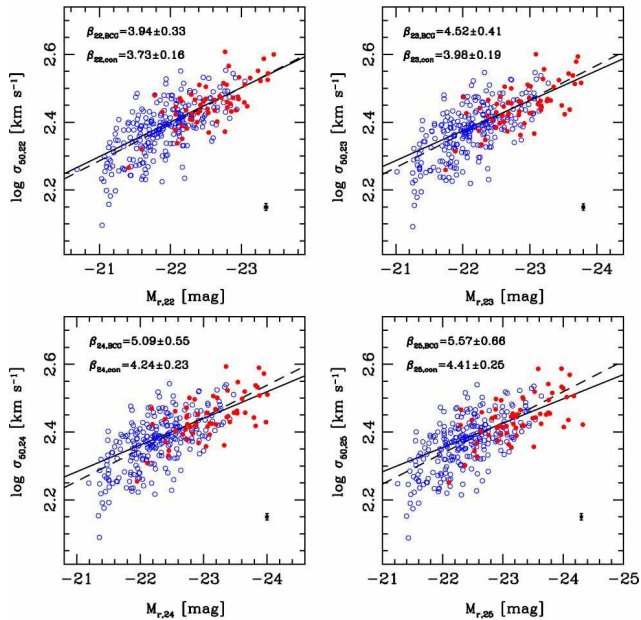}
\caption{ The Faber-Jackson relation for the 59/85 BCGs and 244
control sample galaxies with spectroscopic information for four
different isophotal limits (22, 23, 24, and 25 $\magsec$)
respectively. The symbols are the same as in
Fig.~\ref{size_L.eps}. The power-law indices ${\beta}$ are shown
at the top of each panel for the BCGs and the control sample. The
median error bar in the velocity dispersion is shown at the bottom
right in each panel. } \label{L_sigma.eps}
\end{figure}

\subsubsection{The Faber-Jackson relation and Fundamental plane}

It is well known that the luminosities and velocity dispersions of
many early type galaxies satisfy the Faber-Jackson relation, $L
\propto \sigma^{\;\beta}$, where $\beta \approx 4$ (Faber \&
Jackson 1976). Given that the SDSS spectra were taken within a
3$\arcsec$ fiber aperture (J$\o$rgensen et al. 1995; Bernardi et
al. 2003a), we correct the velocity dispersions for BCGs and the
control sample following von der Linden et al. (2007, see their
equation 4). The uncorrected value of $\sigma$ for each BCG is
shown in Column 5 of Table 2.

Fig.~\ref{L_sigma.eps} shows the $L-\sigma$ relation for 59 out of
85 BCGs and the control sample with SDSS spectral information,
measured to four different isophotal limits, 22, 23, 24, and 25
$\magsec$. Clearly at a given surface brightness limit, the slope
for BCGs in the Fabor-Jackson relation is steeper than that of the
bulk of early type galaxies.  The index $\beta$ for both the BCG
and control samples increases as the isophotal limits become
deeper. The $\beta$ value for the control sample is around 4,
consistent with the canonical value for the Faber-Jackson
relation. However, this is not the case for BCGs: only when the
isophotal limit is 22 $\magsec$, corresponding to the inner
region of BCGs, $\beta$ is about 4. For deeper isophotal limits,
the $\beta$ values for BCGs are significantly larger,
approaching 5.6 for a surface brightness limit of $25 \magsec$.
von der Linden (2007) found $\beta=5.32 \pm 0.37$ for their BCGs
at an isophotal limit of $23\magsec$, which is within $2\sigma$ of
our value ($\beta=4.52 \pm 0.41$). For their control samples, they
found $\beta=3.93 \pm 0.21$, within $1\sigma$ of our result
($\beta=3.98 \pm 0.19$). Lauer found $\beta=6.5 \pm 1.3$ for their
core galaxies with $M_{V}<-21$, which is significantly larger than
those of values mentioned above. Note that the sample Lauer et al.
(2007) used is not a pure BCG sample, although most of their core galaxies with
$M_{V}<-22$ are BCGs. 

\begin{figure}
\centering
\includegraphics[width = 8.4cm]{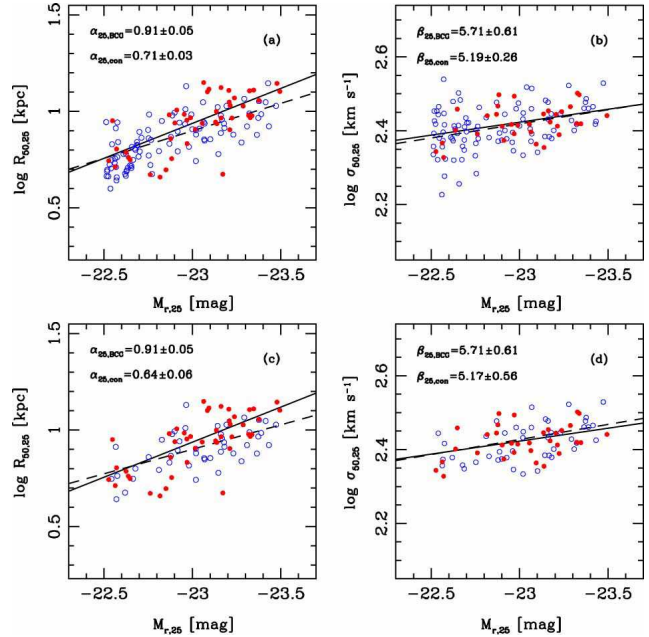}
\caption{ 
The size-luminosity relation (left) and Faber-Jackson
relation (right) for the BCG and control sample galaxies at $-23.5
\leq M_{r,25} \leq -22.5$. The top panels use all 94 control
galaxies and 45 BCGs in this luminosity range. The bottom panels are
the results for one realisation for 45 BCGs and randomly selected 45 control galaxies with the same luminosity distribution.
The symbols are the same as in
Fig.~\ref{size_L.eps}. The power-law indices are shown at the top
of each panel for each sample. 
}
\label{scaling_part.eps}
\end{figure}

The steeper power-law indices
in the size-luminosity and (to a less extent) in the Faber-Jackson relations may arise because
BCGs are brighter than the control sample or because there are intrinsic
differences between BCGs and the bulk of elliptical galaxies.
In order to differentiate these two possibilities, we repeat the above
analysis for our BCG and control samples in the same magnitude
range,  $-23.5 \leq M_{r,25} \leq -22.5$ (see \S\ref{sec:sample}). 
Fig.~\ref{scaling_part.eps} shows  the size-luminosity (left) and Faber-Jackson (right) relations.
The top panels use all the 94 control-sample galaxies and 
45 BCGs within the luminosity range. Since Fig. \ref{redz_col.eps}
shows the control sample is still fainter on average than the BCG
sample within this narrow range, we have randomly selected  
45 out of the 94 control-sample galaxies so they match
well with the luminosity distribution of the 45 BCGs. The statistical
results for different realisations are similar; the bottom panels
shows the results for one realisation. From this figure, we clearly see that 
the power-law indices for the size-luminosity relations 
for BCGs are steeper than those for bright non-BCG elliptical
galaxies; the power-law indices for the Faber-Jackson relations are also steeper, but only at $\sim 1\sigma$ levels,
and so are not statistically significant.
To summarise, the data are consistent with a scenario where
the cluster environment plays an important role on the BCG formation.
However, the numbers of galaxies in our control and BCG samples are
still too small to differentiate the two possibilities mentioned above.


\begin{figure}
\centering
\includegraphics[width = 8.4cm]{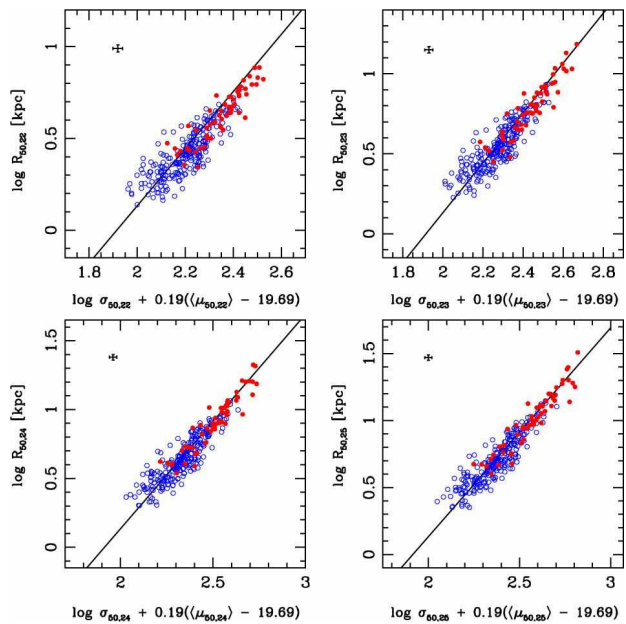}
\caption{ Fundamental planes for the 59 BCGs with spectra (dots)
and 244 control sample (open circles) for four different isophotal
limits respectively. The solid line in each panel is the
fundamental plane determined by Bernardi et al. (2003b) using a
uniform sample of $\sim$9000 early type galaxies. The median error
bars are shown at the top left. } \label{FP.eps}
\end{figure}

Fig.~\ref{FP.eps} shows the fundamental plane for our BCGs and the
control sample. While they appear to have different power-law
slopes in the Faber-Jackson relation, both BCGs and the control
sample follow roughly the same fundamental plane, although the
scatters in the BCGs appear smaller. At bright
isophotal limits, $\mu=22$, and $23\magsec$, the fundamental
planes for BCGs and the control samples appear to differ somewhat.
These results are in good agreement with Bernardi et al. (2007).

\section{SUMMARY} \label{sec:summary}

In this paper we have studied the properties of 85 BCGs carefully selected
from the C4 cluster catalog (Miller et al. 2005). To address the
problems of the SDSS pipeline {\tt PHOTO} in crowded fields and
for large galaxies, we have performed our own photometry for these
BCGs, as well as for a control sample of galaxies in similar
redshift and apparent magnitude ranges. Particular attention was
paid to accurate extraction of the sky background in clusters using a
method developed for high-precision photometry for large, nearby
spiral galaxies (Zheng et al. 1999; Wu et al. 2002, see
\S\ref{sec:data}). It came as somewhat a surprise to us that the
simple recipe of von der Linden et al. (2007) gives surface
brightness profiles similar to ours, and hence for many
statistical purposes their simple recipe can be adopted as a more
efficient alternative.

We analysed the surface brightness properties of our BCGs using
the $\eta(r)$ profiles (see eq. \ref{eq:eta}). We demonstrate that
its gradient, $\gamma(r)\equiv d\eta(r)/d\log(r)$, can be used as
an indicator of extended envelopes of cD galaxies (Patel et al.
2006). Furthermore, we find that the depth of valley, $\Dmin$ in
the $\gamma(r)$ profiles can be used to quantitatively measure the
properties of the plateaus in the surface brightness profiles.
Generally, the deeper the valleys are (corresponding to a larger
value of $\Dmin$), the flatter and wider the plateaus are in the
$\eta(r)$ profiles.
However, the $\Dmin$ parameter appears to vary in a continuous way
with the galaxy parameters, such as their size and luminosity.
While there is a clear trend that more luminous and larger BCGs
have larger stellar halos. the continuous variation makes it
difficult to un-ambiguously classify cD or non-cD galaxies based
on the surface brightness profiles alone.

We measured the photometric properties of galaxies to four
different isophotal limits, $22, 23, 24$, and $25\magsec$, and
used these to investigate the scaling laws and their dependence on
these limits. The latter is of particular relevance as some
previous investigations derived scaling relations for BCGs using
different samples and different isophotal limits (e.g., the study
of von der Linden et al. 2007 used an isophotal limit of
23$\magsec$). We find that the size-luminosity relation and
Fabor-Jackson ($L-\sigma$) relations of BCGs are consistent with being
steeper than those of the bulk of early type galaxies for a control sample
selected within the same absolute magnitude range ($-23.5<M_{r,
25}<-22.5$), although the latter is not statistically significant due
to the limited sample sizes.

Furthermore, as the photometric limit becomes deeper, the size-luminosity relation
for BCGs and the bulk of the ellipticals becomes steeper. However,
there are already (small) differences between BCGs and regular
ellipticals at $22\magsec$, a surface brightness that probes the
inner parts of galaxies. This, together with the relatively small
difference in the scaling relations between $24\magsec$ and
$25\magsec$, suggests that there may be intrinsic differences in
the dynamic structures between BCGs and the bulk of early type
galaxies (Bernardi et al. 2007; Lauer et al. 2007). One
possibility may be that the amount of dark matter in the central
parts of BCGs and regular ellipticals is different (von der Linden
et al. 2007), or perhaps the relative importance of an-isotropy
and rotation may be different. In this context, it is interesting
to note that we find a large fraction of very luminous early type
galaxies with disky isophotes.
This appears to somewhat contradict with the usual expectation that the
most luminous galaxies should show boxy isophotes. Also notice
that it is difficult to measure high-redshift galaxies to very
faint surface brightness due to the cosmological dimming. Thus a proper comparison between local and
high-redshift samples will need to explicitly account for the
surface brightness dependence we found in the paper.

In this work we uncovered four strong dry merger candidate BCGs (see
Fig.~\ref{merger_sbp_eta.eps}), three of which have two close nuclei
within $10$\,kpc and with nearly identical radial velocities. We also found
another 10 BCGs in our sample with possible dry merger 
signatures (see Fig.~\ref{example_merger.eps}), as advocated by Bell et al. (2006).
This strongly suggests merging at the centre of clusters is an important physical
process that affects a variety of properties of BCGs, both in photometry and
kinematics. An analysis of the statistics for the fraction of
dry mergers in BCGs is ongoing and the results will be reported elsewhere.

\section*{Acknowledgements}

We thank C. N. Hao, Simon White for discussions, and in particular
A. von der Linden for sharing data, helpful discussions and comments on a draft. 
We acknowledge Dr. Tod Lauer, the referee, for a constructive report that improved the paper. 
This project is supported by the NSF of China 10333060, 10273012,
10640430201, 10773014 and 973 programs No. 2007CB815405 and 2007CB815406. SM acknowledges 
the Chinese Academy of Sciences and the Alexander von Humboldt
Foundation.
Funding for the creation and distribution of the SDSS Archive has
been provided by the Alfred P. Sloan Foundation, the Participating
Institutions, the National Aeronautics and Space Administration,
the National Science Foundation, the U.S. Department of Energy,
the Japanese Monbukagakusho, and the Max Planck Society. The SDSS
Web site is http://www.sdss.org/. The SDSS is managed by the
Astrophysical Research Consortium (ARC) for the Participating
Institutions. The Participating Institutions are The University of
Chicago, Fermilab, the Institute for Advanced Study, the Japan
Participation Group, The Johns Hopkins University, the Korean
Scientist Group, Los Alamos National Laboratory, the
Max-Planck-Institute for Astronomy (MPIA), the
Max-Planck-Institute for Astrophysics (MPA), New Mexico State
University, University of Pittsburgh, Princeton University, the
United States Naval Observatory, and the University of Washington.

\label{lastpage}


\begin{thebibliography}{99}

\bibitem[\protect\citeauthoryear{Bell et al.}{2006}]{Bell06} Bell E. F. et al., 2006, ApJ, 640, 241
\bibitem[\protect\citeauthoryear{Bender}{1988}]{Ben88} Bender R., 1988, A\&A, 193, L7
\bibitem[\protect\citeauthoryear{Bender et al.}{1988}]{Bender88} Bender R., D\"obereiner S., M\"ollenhoff C., 1988, A\&AS, 74, 385
\bibitem[\protect\citeauthoryear{Bender et al.}{1989}]{Bender89} Bender R., Surma P., Doebereiner S., Moellenhoff C., Madejsky R., 1989, A\&A, 217, 35
\bibitem[\protect\citeauthoryear{Bernardi et al.}{2003a}]{Bernardi03a} Bernardi M. et al., 2003a, AJ, 125, 1817
\bibitem[\protect\citeauthoryear{Bernardi et al.}{2003b}]{Bernardi03b} Bernardi M. et al., 2003b, AJ, 125, 1866
\bibitem[\protect\citeauthoryear{Bernardi et al.}{2007}]{Bernardi07} Bernardi M., Hyde J. B., Sheth R. K., Miller C. J. Nichol R. C., 2007, AJ, 133, 1741
\bibitem[\protect\citeauthoryear{Bertin \& Arnouts}{1996}]{Bertin96} Bertin E., Arnouts S., 1996, A\&AS, 117, 393
\bibitem[\protect\citeauthoryear{Binney \& Merrifield}{1998}]{Binney98} Binney J., Merrifield M., 1998, Galactic Astronomy (Princeton: Princeton Univ. Press)
\bibitem[\protect\citeauthoryear{Blanton et al.}{2003a}]{Blanton03a} Blanton M. R. et al., 2003a, ApJ, 592, 819
\bibitem[\protect\citeauthoryear{Blanton et al.}{2003b}]{Blanton03b} Blanton M. R. et al., 2003b, ApJ, 594, 186
\bibitem[\protect\citeauthoryear{Blanton et al.}{2005}]{Blanton05} Blanton M. R. et al., 2005, AJ, 129, 2562
\bibitem[\protect\citeauthoryear{Blanton \& Roweis}{2007}]{Blanton07} Blanton M. R., Roweis S., 2007, AJ, 133, 734
\bibitem[\protect\citeauthoryear{Brough et al.}{2005}]{Brough05} Brough S., Collins C. A., Burke D. J., Lynam P. D., Mann R. G., 2005, MNRAS, 364, 1354
\bibitem[\protect\citeauthoryear{Cowie \& Binney}{1977}]{Cowie77} Cowie L. L., Binney J., 1977, ApJ, 215, 723
\bibitem[\protect\citeauthoryear{Davies et al.}{1983}]{Davies83} Davies R. L., Efstathiou G., Fall S. M., Illingworth G., Schechter P. L., 1983, ApJ, 266, 41
\bibitem[\protect\citeauthoryear{De Lucia \& Blaizot}{2007}]{DeLucia07} De Lucia G., Blaizot J., 2007, MNRAS, 375, 2
\bibitem[\protect\citeauthoryear{Desroches et al.}{2007}]{Desroches07} Desroches Louis-Benoit, Quataert E., Ma Chung-Pei, West A. A., 2007, MNRAS, 377, 402
\bibitem[\protect\citeauthoryear{de Vaucouleurs}{1948}]{deVau48} de Vaucouleurs G., 1948, Ann. Astrophys., 11, 247
\bibitem[\protect\citeauthoryear{Faber et al.}{1997}]{Faber97} Faber S. M. et al., 1997, AJ, 114, 1771
\bibitem[\protect\citeauthoryear{Faber et al.}{1976}]{Faber76} Faber S. M., Jackson R. E., 1976, ApJ, 204, 668
\bibitem[\protect\citeauthoryear{Fabian}{1994}]{Fabian94} Fabian A. C., 1994, ARA\&A, 32, 277
\bibitem[\protect\citeauthoryear{Fabian \& Nulsen}{1977}]{Fabian77} Fabian A. C., Nulsen P. E. J., 1977, MNRAS, 180, 479
\bibitem[\protect\citeauthoryear{Fan et al.}{1996}]{Fan96} Fan X. et al., 1996, AJ, 112, 628
\bibitem[\protect\citeauthoryear{Gallagher \& Ostriker}{1972}]{Gallagher72} Gallagher J. S., Ostriker J. P., 1972, AJ, 77, 288
\bibitem[\protect\citeauthoryear{Gao et al.}{2004}]{Gao04} Gao L., Loeb A., Peebles P. J. E., White S. D. M., Jenkins A., 2004, ApJ, 614, 17
\bibitem[\protect\citeauthoryear{Gonzalez et al.}{2005}]{Gonzalez05} Gonzalez A. H., Zabludoff A. I., Zaritsky D., 2005, ApJ, 618, 195
\bibitem[\protect\citeauthoryear{Graham et al.}{1996}]{Graham96} Graham A., Lauer T. R., Colless M., Postman M., 1996, ApJ, 465, 534
\bibitem[\protect\citeauthoryear{Hao et al.}{2006}]{Hao06} Hao C. N., Mao S., Deng Z. G., Xia X. Y., Wu H., 2006, MNRAS, 370, 1339; erratum, MNRAS, 373, 1264
\bibitem[\protect\citeauthoryear{Jones \& Forman}{1984}]{Jones84} Jones C., Forman W., 1984, ApJ, 276, 38
\bibitem[\protect\citeauthoryear{Jorgensen et al.}{1995}]{Jorgensen95} J$\o$rgensen I., Franx M., Kj$\ae$rgaard P, 1995, MNRAS, 276, 1341
\bibitem[\protect\citeauthoryear{Kjaergaard et al.}{1993}]{Kjergaard93} Kj$\ae$rgaard P., J$\o$rgensen I., Moles M., 1993, ApJ, 418, 617
\bibitem[\protect\citeauthoryear{Koester et al.}{2007}]{Koester07} Koester B. P. et al., 2007, ApJ, 660, 239
\bibitem[\protect\citeauthoryear{Krick et al.}{2006}]{Krick06} Krick J. E., Bernstein R. A., Pimbblet K. A., 2006, AJ, 131, 168
\bibitem[\protect\citeauthoryear{Laine et al.}{2003}]{Laine03} Laine S., van der Marel R. P., Lauer T. R., Postman M., O'Dea C. P., Owen F. N., 2003, AJ, 125, 478
\bibitem[\protect\citeauthoryear{Lauer}{1985}]{Lauer85} Lauer T. R., 1985, MNRAS, 216, 429
\bibitem[\protect\citeauthoryear{Lauer}{1988}]{Lauer88} Lauer T. R., 1988, ApJ, 325, 49 
\bibitem[\protect\citeauthoryear{Lauer et al.}{2005}]{Lauer05} Lauer T. R. et al., 2005, AJ, 129, 2138
\bibitem[\protect\citeauthoryear{Lauer et al.}{2007}]{Lauer07} Lauer T. R. et al., 2007, ApJ, 662, 808
\bibitem[\protect\citeauthoryear{Lin}{2004}]{Lin04} Lin Y.-T., Mohr J. J., 2004, ApJ, 617, 879
\bibitem[\protect\citeauthoryear{Lugger}{1984}]{Lugger84} Lugger P. M., 1984, ApJ, 286, 106
\bibitem[\protect\citeauthoryear{Matthews}{1964}]{Matthews64} Matthews T. A., Morgan W. W., Schmidt M., 1964, ApJ, 140, 35
\bibitem[\protect\citeauthoryear{Merritt}{1985}]{Merritt85} Merritt D., 1985, ApJ, 289, 18
\bibitem[\protect\citeauthoryear{Michard}{2002}]{Michard02} Michard R., 2002, A\&A, 384, 763
\bibitem[\protect\citeauthoryear{Miller et al.}{2005}]{Miller05} Miller C. J. et al., 2005, AJ, 130, 968
\bibitem[\protect\citeauthoryear{Nieto et al.}{1988}]{Nieto88} Nieto J.-L., Capaccioli M., Held V. E., 1988, A\&A, 195, L1
\bibitem[\protect\citeauthoryear{Oegerle}{1991}]{Oegerle91} Oegerle W. R., Hoessel J. G., 1991, ApJ, 375, 15
\bibitem[\protect\citeauthoryear{Oemler}{1973}]{Oemler73} Oemler A., Jr., 1973, ApJ, 180, 11
\bibitem[\protect\citeauthoryear{Oemler}{1976}]{Oemler76} Oemler A., Jr., 1976, ApJ, 209, 693 
\bibitem[\protect\citeauthoryear{Ostriker \& Hausman}{1977}]{Ostriker77} Ostriker J. P., Hausman M. A., 1977, ApJ, 217, L125
\bibitem[\protect\citeauthoryear{Ostriker \& Tremaine}{1975}]{Ostriker75} Ostriker J. P., Tremaine S. D., 1975, ApJ, 202, L113
\bibitem[\protect\citeauthoryear{Patel et al.}{2006}]{Patel06} Patel P., Maddox S., Pearce F.~R., Arag\'on-Salamanca A., Conway E., 2006, MNRAS, 370, 851
\bibitem[\protect\citeauthoryear{Petrosian}{1976}]{Petrosian} Petrosian V., 1976, ApJ, 209, L1
\bibitem[\protect\citeauthoryear{Postman \& Lauer}{1995}]{PL} Postman M., Lauer T. R., 1995, ApJ, 440, 28
\bibitem[\protect\citeauthoryear{Rest et al.}{2001}]{Rest01} Rest A., van den Bosch F. C., Jaffe W., Tran H., Tsvetanov Z., Ford H. C., Davies J., Schafer J., 2001, AJ, 121, 2431
\bibitem[\protect\citeauthoryear{Richstone}{1975}]{Richstone75} Richstone D., 1975, ApJ, 200, 535
\bibitem[\protect\citeauthoryear{Richstone}{1976}]{Richstone76} Richstone D., 1976, ApJ, 204, 642
\bibitem[\protect\citeauthoryear{Schombert}{1986}]{Schombert86} Schombert J. M., 1986, ApJS, 60, 603
\bibitem[\protect\citeauthoryear{Schombert}{1987}]{Schombert87} Schombert J. M., 1987, ApJS, 64, 643
\bibitem[\protect\citeauthoryear{Schombert}{1988}]{Schombert88} Schombert J. M., 1988, ApJ, 328, 475
\bibitem[\protect\citeauthoryear{Sersic}{1968}]{Sersic68} S$\acute{e}$rsic J. L., 1968, Atlas de Galaxies Australes (Cordoba: Observatories Astronomica)
\bibitem[\protect\citeauthoryear{Shen et al.}{2003}]{Shen03} Shen S. et al., 2003, MNRAS, 343, 978
\bibitem[\protect\citeauthoryear{Shimasaku et al.}{2001}]{Shi01} Shimasaku K. et al., 2001, AJ, 122, 1238
\bibitem[\protect\citeauthoryear{Smith et al.}{2005}]{Smith05} Smith G. P., Kneib J.-P., Smail I., Mazzotta P., Ebeling H., Czoske O., 2005, MNRAS, 359, 417
\bibitem[\protect\citeauthoryear{Stoughton et al.}{2002}]{Stoughton02} Stoughton C. et al., 2002, AJ, 123, 485
\bibitem[\protect\citeauthoryear{Strateva et al.}{2001}]{Str01} Strateva I. et al., 2001, AJ, 122, 1861
\bibitem[\protect\citeauthoryear{Tran et al.}{2005}]{Tran05} Tran K.-V., van Dokkum P. G., Franx M., Illingworth G. D., Kelson D. D., Schreiber N. M. F., 2005, ApJ, 627, L25
\bibitem[\protect\citeauthoryear{van den Bosch et al.}{1994}]{Bosch94} van den Bosch F. C., Ferrarese L., Jaffe W., Ford H. C., O'Connell R. W., 1994, AJ, 108, 1579
\bibitem[\protect\citeauthoryear{van Dokkum}{2005}]{Dokkum05} van Dokkum P. G., 2005, AJ, 130, 2647
\bibitem[\protect\citeauthoryear{von der Linden et al.}{2007}]{Linden07} von der Linden A. et al., 2007, MNRAS, 379, 867
\bibitem[\protect\citeauthoryear{White}{1976}]{White76} White S. D. M., 1976, MNRAS, 174, 19
\bibitem[\protect\citeauthoryear{White}{2005}]{White05} White S. D. M. et al., 2005, AAP, 444, 365
\bibitem[\protect\citeauthoryear{Wu et al.}{2002}]{Wu02} Wu H. et al., 2002, AJ, 123, 1364
\bibitem[\protect\citeauthoryear{Wu et al.}{2005}]{Wu05} Wu H., Shao Z. Y., Mo H. J., Xia X. Y., Deng Z. G., 2005, ApJ, 622, 244
\bibitem[\protect\citeauthoryear{Zebetti et al.}{2005}]{Zebetti05} Zibetti S., White S.~D.~M., Schneider D.~P., Brinkmann J., 2005, MNRAS, 358, 949
\bibitem[\protect\citeauthoryear{Zheng et al.}{1999}]{Zheng99} Zheng Z. Y. et al., 1999, AJ, 117, 2757

\end{thebibliography}
\end{document}